%% file: main.tex
\documentclass[12pt,a4paper]{article}

\usepackage{ifthen} 
\newboolean{articletitles}
\setboolean{articletitles}{true} 

\newboolean{uprightparticles}
\setboolean{uprightparticles}{false} 

\newboolean{inbibliography}
\setboolean{inbibliography}{false} 

\input{preamble}
\usepackage{longtable} 

\newcommand{\FE}{\begin{small}FE\end{small}\,}
\newcommand{\ASDBLR}{\begin{small}ASDBLR\end{small}\,}

\begin{document}

\renewcommand{\thefootnote}{\fnsymbol{footnote}}
\setcounter{footnote}{1}


\begin{titlepage}
\pagenumbering{roman}

\vspace*{-1.5cm}
\centerline{\large EUROPEAN ORGANIZATION FOR NUCLEAR RESEARCH (CERN)}
\vspace*{1.5cm}
\hspace*{-0.5cm}

\begin{tabular*}{\linewidth}{lc@{\extracolsep{\fill}}r}
\vspace*{-1.2cm}\mbox{\!\!\!\includegraphics[width=.12\textwidth]{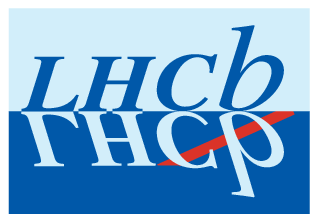}} & &%
\\
 & & LHCb-DP-2013-003 \\  
 & & 7 January 2013 \\ 
 & & \\
\end{tabular*}

\vspace*{2.0cm}

{\bf\boldmath\huge
\begin{center}
  Performance of the \lhcb Outer Tracker
\end{center}
}

\vspace*{0.0cm}

\begin{center}
The LHCb Outer Tracker group
\input{OT_authorlist.tex}

\end{center}

\vspace{\fill}

\begin{abstract}
  \noindent
The LHCb Outer Tracker is a gaseous detector covering an area of $5\times
6$\,m$^2$ with 12 double layers of straw tubes.  The detector with its services
are described together with the commissioning and calibration procedures.  Based
on data of the first LHC running period from 2010 to 2012, the performance of
the readout electronics and the single hit resolution and efficiency are
presented.

The efficiency to detect a hit in the central half of the straw is estimated to be
99.2\%, and the position resolution is determined to be approximately
200\,$\mum$.  The Outer Tracker received a dose in the hottest region
corresponding to 0.12\,C/cm, and no signs of gain deterioration or other ageing
effects are observed.
\end{abstract}

\vspace*{0.0cm}

\begin{center}
  Published in JINST
\end{center}

\vspace{\fill}

{\footnotesize 
\centerline{\copyright~CERN on behalf of the \lhcb collaboration, 
license \href{http://creativecommons.org/licenses/by/3.0/}{CC-BY-3.0}.}}
\vspace*{2mm}

\end{titlepage}


\newpage
\setcounter{page}{2}
\mbox{~}
\newpage

\cleardoublepage


\renewcommand{\thefootnote}{\arabic{footnote}}
\setcounter{footnote}{0}

\tableofcontents


\pagestyle{plain} 
\setcounter{page}{1}
\pagenumbering{arabic}


\input{content}

\addcontentsline{toc}{section}{References}
\setboolean{inbibliography}{true}
\bibliographystyle{LHCb}
\bibliography{main,main-ot}

\end{document}

%% file: preamble.tex
\usepackage{microtype}
\usepackage{lineno}  
\usepackage{xspace} 
\usepackage{caption} 

\usepackage{graphicx}  
\usepackage{color}
\usepackage{colortbl}
\graphicspath{{./figs/}} 

\usepackage{amsmath} 
\usepackage{amssymb}
\usepackage{amsfonts}
\usepackage{upgreek} 

\newcommand*\patchAmsMathEnvironmentForLineno[1]{%
\expandafter\let\csname old#1\expandafter\endcsname\csname #1\endcsname
\expandafter\let\csname oldend#1\expandafter\endcsname\csname
end#1\endcsname
 \renewenvironment{#1}%
   {\linenomath\csname old#1\endcsname}%
   {\csname oldend#1\endcsname\endlinenomath}%
}
\newcommand*\patchBothAmsMathEnvironmentsForLineno[1]{%
  \patchAmsMathEnvironmentForLineno{#1}%
  \patchAmsMathEnvironmentForLineno{#1*}%
}
\AtBeginDocument{%
\patchBothAmsMathEnvironmentsForLineno{equation}%
\patchBothAmsMathEnvironmentsForLineno{align}%
\patchBothAmsMathEnvironmentsForLineno{flalign}%
\patchBothAmsMathEnvironmentsForLineno{alignat}%
\patchBothAmsMathEnvironmentsForLineno{gather}%
\patchBothAmsMathEnvironmentsForLineno{multline}%
}

\usepackage{hyperref}    
\usepackage[all]{hypcap} 

\def\lhcb {LHCb\xspace}
\def\mum  {\ensuremath{\,\upmu\rm m}\xspace}
\newcommand{\ie}{\mbox{\itshape i.e.}}

\usepackage{cite} 
\usepackage{mciteplus}

%% file: OT_authorlist.tex


\begin{flushleft}
\center
\small
R. Arink$^{       1}$,
S. Bachmann$^{	  2}$,
Y. Bagaturia$^{	  2}$,
H. Band$^{	  1}$,
Th. Bauer$^{	  1}$,
A. Berkien$^{	  1}$,
Ch. F\"arber$^{	  2}$,
A. Bien$^{	  2}$,
J. Blouw$^{	  2}$,
L. Ceelie$^{	  1}$,
V. Coco$^{	  1}$,
M. Deckenhoff$^{  3}$,
Z. Deng$^{	  7}$,
F. Dettori$^{	  1}$,
D. van Eijk$^{	  1}$,
R. Ekelhof$^{	  3}$,
E. Gersabeck$^{	  2}$,
L. Grillo$^{	  2}$,
W.D. Hulsbergen$^{1}$,
T.M. Karbach$^{	  3,4}$,
R. Koopman$^{	  1}$,
A. Kozlinskiy$^{  1}$,
Ch. Langenbruch$^{2}$,
V. Lavrentyev$^{  1}$,
Ch. Linn$^{	  2}$,
M. Merk$^{	  1}$,
J. Merkel$^{	  3}$,
M. Meissner$^{	  2}$,
J. Michalowski$^{ 5}$,
P. Morawski$^{	  5}$,
A. Nawrot$^{	  6}$,
M. Nedos$^{	  3}$,
A. Pellegrino$^{  1}$,
G. Polok$^{	  5}$,
O. van Petten$^{  1}$,
J. R\"ovekamp$^{  1}$,
F. Schimmel$^{	  1}$,
H. Schuylenburg$^{1}$,
R. Schwemmer$^{	  2,4}$,
P. Seyfert$^{	  2}$,
N. Serra$^{	  1}$,
T. Sluijk$^{	  1}$,
B. Spaan$^{	  3}$,
J. Spelt$^{	  1}$,
B. Storaci$^{	  1}$,
M. Szczekowski$^{ 6}$,
S. Swientek$^{	  3}$,
S. Tolk$^{	  1}$,
N. Tuning$^{	  1}$,
U. Uwer$^{	  2}$,
D. Wiedner$^{	  2}$,
M. Witek$^{	  5}$,
M. Zeng$^{	  7}$,
A. Zwart$^{	  1}$.\bigskip\newline{\it
\footnotesize
$ ^{1}$Nikhef, Amsterdam, The Netherlands\\
$ ^{2}$Physikalisches Institut, Heidelberg, Germany\\
$ ^{3}$Technische Universit\"at Dortmund, Germany\\
$ ^{4}$CERN, Geneva, Switzerland\\
$ ^{5}$H. Niewodniczanski Institute of Nuclear Physics, Cracow, Poland\\
$ ^{6}$A. Soltan Institute for Nuclear Studies, Warsaw, Poland\\
$ ^{7}$Tsinghua University, Beijing, China\\
}
\end{flushleft}

%% file: content.tex
\clearpage
\section{Introduction}
\label{sec:introduction}
The LHCb detector~\cite{Alves:2008zz} is a single-arm forward spectrometer
covering the pseudo-rapidity range $2<\eta <5$, designed for the study of
particles containing $b$ or $c$ quarks. The detector includes a high-precision
tracking system consisting of a silicon-strip vertex detector surrounding the
$pp$ interaction region, a large-area silicon-strip detector located upstream of
a dipole magnet with a bending power of about 4~Tm and three tracking stations
located downstream. The area close to the beamline is covered by silicon-strip
detectors, whereas the large area at more central rapidity is covered by the
Outer Tracker (OT) straw-tube detector.

\begin{figure}[!t]
\begin{center}
    \begin{picture}(250,300)(0,0)
    \put(-105,205){\includegraphics[bb=0 0 1982 875,scale=0.12]{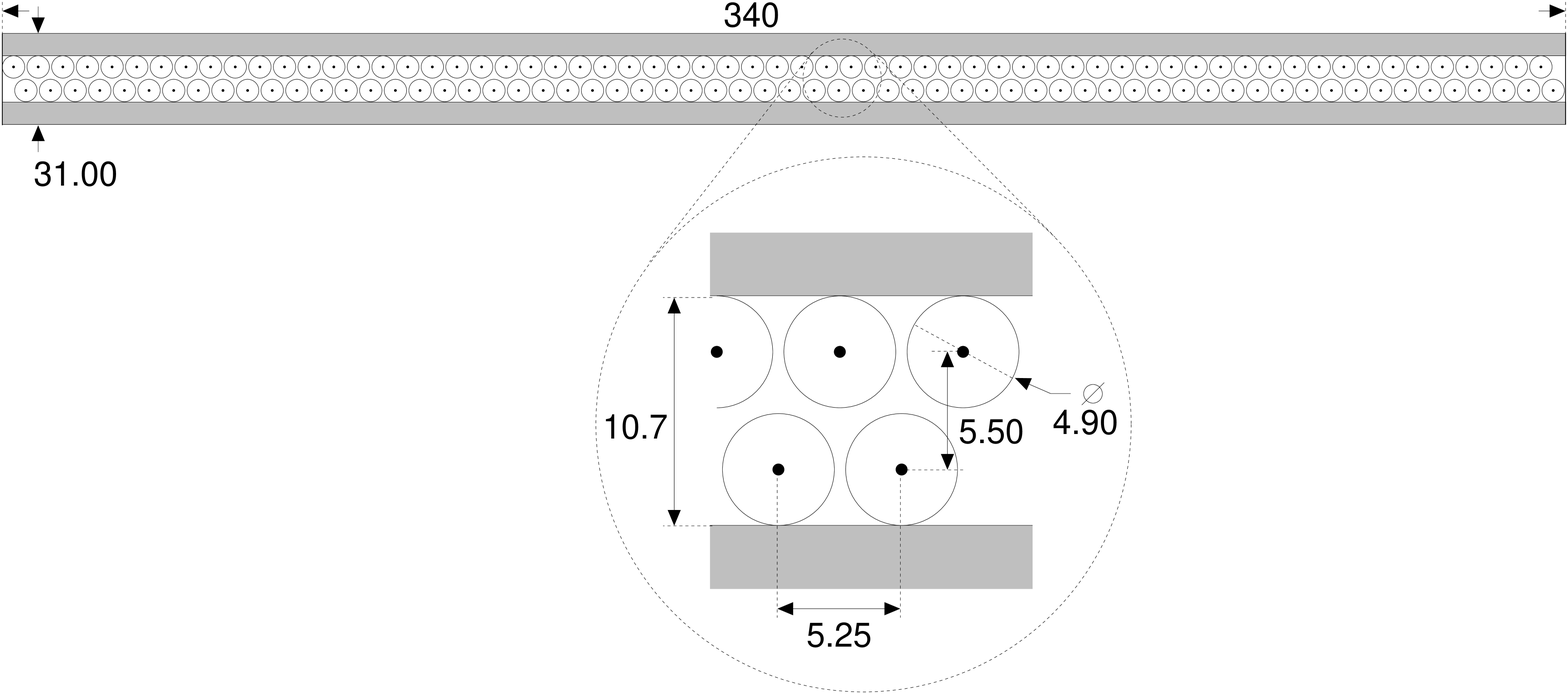}}
    \put(45, 10){\includegraphics[bb=60 127 469 469,scale=0.75]{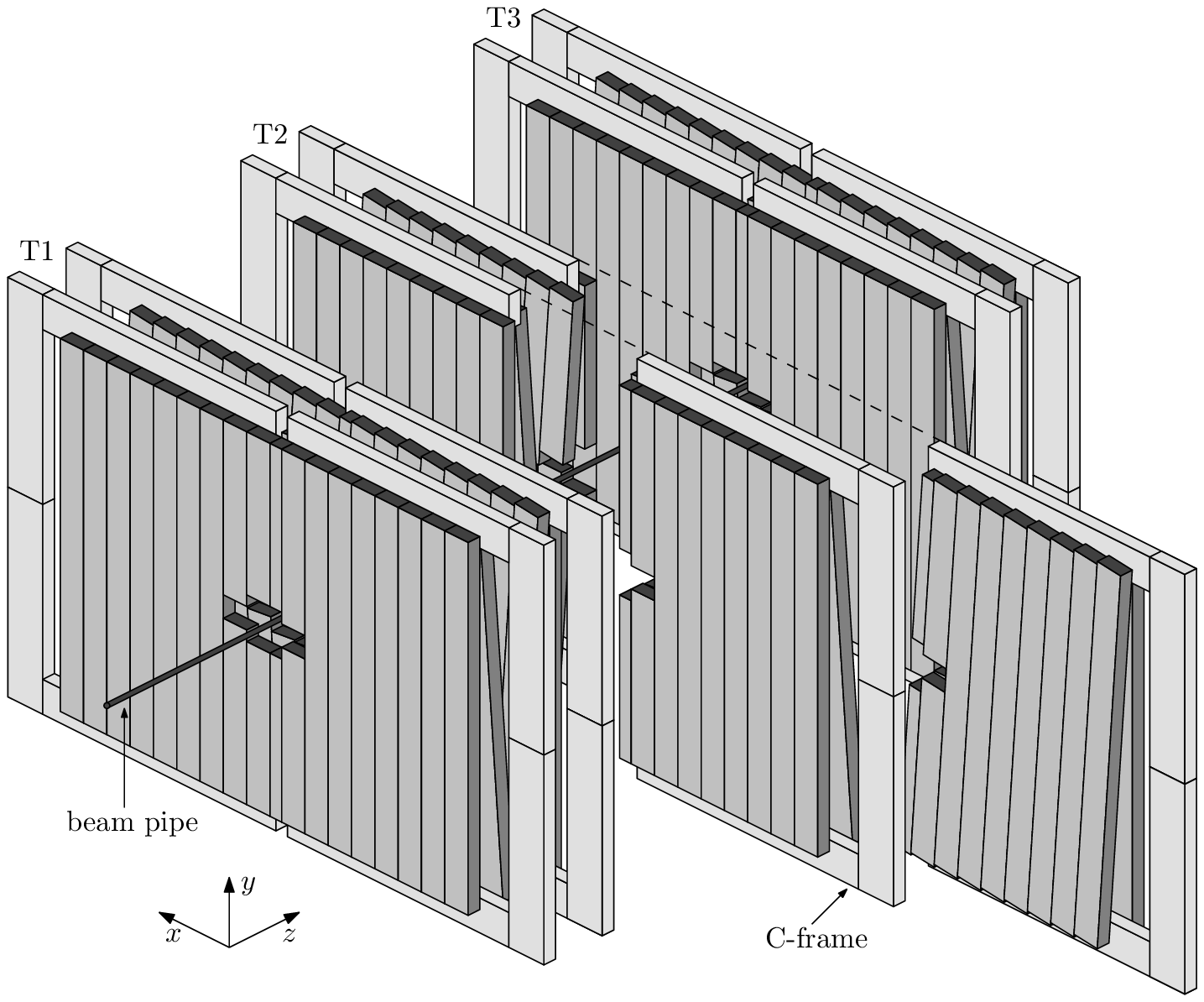}}
    \put(-100,200){(a)}
    \put(30,10){(b)}
    \end{picture}
    \caption[Outer Tracker module cross section]{\small 
    (a) Module cross section.
    (b) Arrangement of OT straw-tube modules in layers and stations.}
    \label{fig:mod}
\end{center}
\end{figure}

Excellent momentum resolution is required for a precise determination of the
invariant mass of the reconstructed $b$-hadrons. For example a mass resolution of
25 MeV/c$^2$ for the decay $B_{s}^0 \rightarrow \mu^+\mu^-$ translates into a
required momentum resolution of $\delta p / p \approx
0.4\%$~\cite{Simioni:2010zz}.  Furthermore, the reconstruction of
high-multiplicity $B$ decays demands a high tracking efficiency and at the same
time a low fraction of wrongly reconstructed tracks.  
To achieve the physics
goals of the LHCb experiment, the OT is required to determine the position of
single hits with a resolution of 200~$\mu$m in the $x$-coordinate~\footnote{ The
LHCb coordinate system is a right-handed coordinate system, with the $z$ axis
pointing along the beam axis, $y$ is the vertical direction, and $x$ is the
horizontal direction. The $xz$ plane is the bending plane of the dipole
magnet.}, while limiting the radiation length to 3\% $X_0$ per station (see
Fig.~\ref{fig:mod}b).  A fast counting gas is needed to keep the occupancy below
10\% at the nominal luminosity of $2\times 10^{32}$\,cm$^{-2}$s$^{-1}$.

The OT is a gaseous straw tube detector~\cite{TDR} and covers an area of
approximately $5\times 6$\,m$^2$ with 12 double layers of straw tubes.  The straw tubes
are 2.4\,m long with 4.9\,mm inner diameter, and are filled with a gas mixture
of Ar/CO$_2$/O$_2$ (70/28.5/1.5) which guarantees a fast drift-time below 50\,ns.
The anode wire is set to +1550\,V and is made of gold plated tungsten of
25\,$\mu$m diameter, whereas the cathode consists of a 40\,$\mu$m thick inner
foil of electrically conducting carbon doped Kapton-XC~\footnote{
Kapton\textregistered~ is a polyimide film developed by DuPont.}  
and a 25\,$\mu$m thick outer foil, consisting of Kapton-XC laminated together
with a 12.5\,$\mu$m thick layer of aluminium.  The straws are glued to sandwich
panels, using Araldite AY103-1~\footnote{Araldite\textregistered~ is a two
component epoxy resin developed by Huntsman.}.  Two panels are sealed with
400\,$\mu$m thick carbon fiber sidewalls, resulting in a gas-tight box enclosing
a stand-alone detector module.  A cross-section of the module layout is shown in
Fig.~\ref{fig:mod}(a).

The modules are composed of two staggered layers (monolayers) of 64 drift tubes
each. In the longest modules (type $F$) the monolayers are split in the middle
into two independent readout sections composed of individual straw tubes. Both
sections are read out from the outer ends. The splitting in two sections is done
at a different position for the two monolayers to avoid insensitive regions in
the middle of the module.  $F$-modules have an active length of 4850\,mm and
contain twice 128 straws, in the upper and the lower half, respectively.  
Short modules (type $S$) have about half the length of $F$-type modules and are
mounted above and below the beampipe.  They contain 128 single drift tubes, and
are read out only from the outer module end.  The inner region not covered by
the OT, $|y|<10(20)$~cm for $|x|<59.7(25.6)$~cm, is instrumented with silicon
strip detectors~\cite{Alves:2008zz}.  One detector layer is built from 14 long
and 8 short modules, see Fig.~\ref{fig:mod}(b).  The complete OT detector
consists of 168 long and 96 short modules and comprises 53,760 single straw-tube
channels.

The detector modules are arranged in three stations.  Each station consists of
four module layers, arranged in an {\it x-u-v-x} geometry: the modules in the
$x$-layers are oriented vertically, whereas those in the $u$ and $v$ layers are
tilted by $+5^o$ and $-5^o$ with respect to the vertical, respectively.
This leads to a total of 24 straw layers positioned along the $z$-axis. 

Each station is split into two halves, retractable on both sides of the beam
line. Each half consists of two independently movable units, known as  
C-frames, see Fig.~\ref{fig:mod}(b).  
The modules are positioned on the C-frames by means of precision
dowel pins.  The C-frames also provide routing for all detector services (gas,
low and high voltage, water cooling, data fibres, slow and fast control).  The
OT C-frames are sustained by a stainless steel structure (OT bridge), equipped
with rails allowing the independent movement of all twelve C-frames. At the top
the C-frames hang on the rails, while at the bottom the C-frames are guided, 
but not supported by the rails, to constrain the movement in the $z$-coordinate.

\begin{figure}[!t]
    \begin{picture}(250,190)(0,0)
    \put(0,25){\includegraphics[bb=0 0 590 405,clip=,scale=0.40]{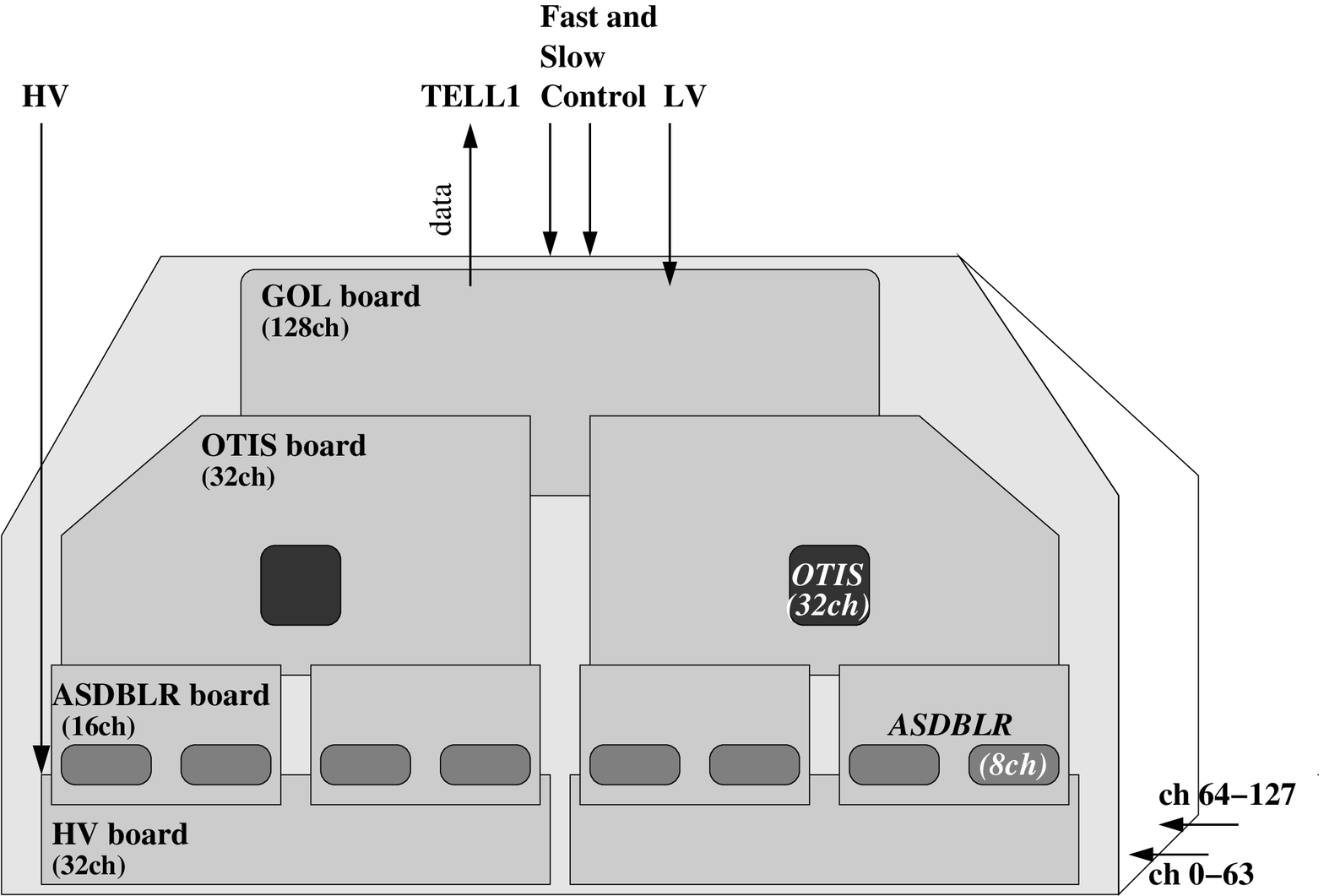}}
    \put(250,25){\includegraphics[bb=0 0 411 309,clip,scale=0.47]{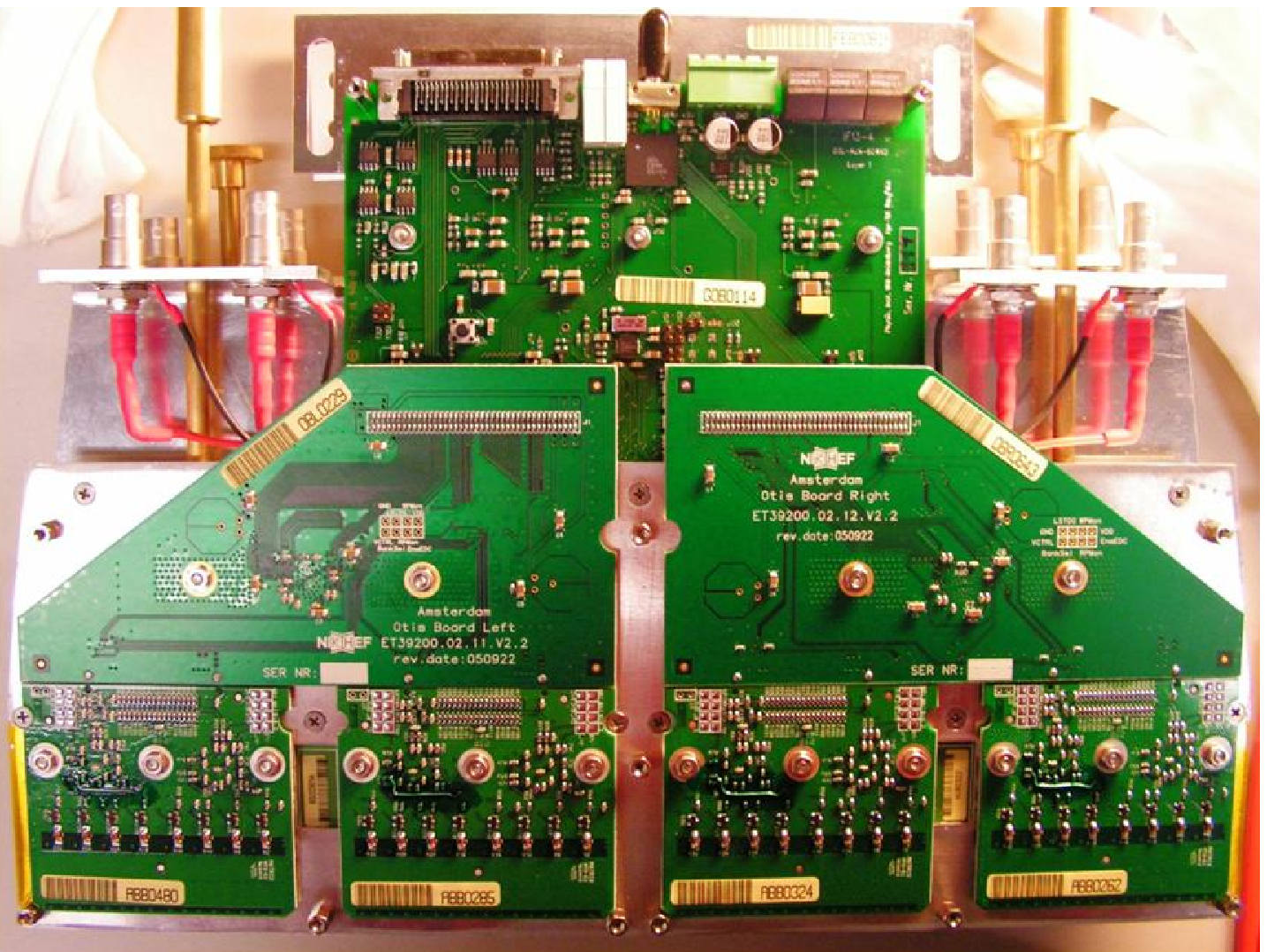}}
    \put(0,0){(a)}
    \put(250,0){(b)}
    \end{picture}
\caption{\small
         (a) Design and (b) photograph of the FE electronics mounted in
 a FE box. Only the boards that read out one monolayer of 64 straws
are visible.
In addition, the HV boards are not visible in the photograph
as they are hidden by the ASDBLR boards.}
\label{fig:fe}
\end{figure}
The front-end (FE) electronics measures the drift-times of the ionization
clusters produced by charged particles traversing the straw-tubes with respect
to the beam crossing (BX) signal~\cite{Berkien:2005zz}.  The drift-times are
digitized for each 25\,ns (the LHC design value for the minimum bunch crossing
interval) and stored in a digital pipeline to await the lowest-level trigger
(L0) decision.  On a positive L0 decision, the digitized data in a window of
75\,ns is transmitted via optical links to TELL1 boards in the LHCb DAQ
system~\cite{Haefeli:2006cv}.

As shown in Fig.~\ref{fig:fe}, the FE electronics has a modular design,
consisting of several interconnected boards housed inside a metallic box (FE
box).  The main components of the OT readout electronics are the high voltage
(HV) board, the ASDBLR amplifier board, the OTIS digitization board, and the GOL
auxiliary (GOL/AUX) board.  Each ASDBLR board hosts two ASDBLR
chips~\cite{Dressnandt:2000fe}.  These are custom-made integrated circuits,
providing the complete analog signal processing chain (amplification, shaping,
baseline restoration, and discrimination) for the straw tube detectors.  The hit
outputs of two ASDBLR boards (32 channels) are connected to one OTIS board,
which hosts one radiation-hard OTIS TDC chip for drift-time
digitization~\cite{Deppe:2008zz,Stange:2005tj}.  The time digitization is done
through the 25\,ns long Delay Locked Loop (DLL) using the 64 delay-stages of the
DLL (64 time bins), giving a step size of about 0.4\,ns.

The drift-time data is stored in a pipeline memory with a depth of 164 events,
corresponding to a latency of 4.1\,$\mu$s.  If a trigger occurs, the
corresponding data words of up to 3 bunch crossings are transferred to a
derandomizing buffer, able to store data from up to 16 consecutive triggers.
Only the first hit in the 75\,ns wide window of a given channel is stored.
Later signals from multiple ionizations or reflections are thus not recorded.
The OTIS boards in a FE box are connected to one GOL/AUX board.  This
board~\cite{GOL} provides the outside connections to the FE box: the power
connection, the interface to the fast-control (beam crossing clock BX, triggers,
resets) and the interface to the slow-control (I$^2$C).

These boxes are mounted at each end of the detector modules.  A FE box is the
smallest independent readout unit of the OT: the digitized data of the 128
channels of one module are sent via an optical link and received by the TELL1
board. High- and low-voltage, as well as fast- and slow-control signals are
connected to each FE box individually.  In total, 432 FE boxes are used to read
out the OT detector.

This paper describes the detector performance in the first LHC running period
from 2010 to 2012, when the LHCb experiment collected data at stable conditions,
corresponding to a typical instantaneous luminosity of about $3.5\,(4.0) \times
10^{32}$\,cm$^{-2}$s$^{-1}$ in 2011 (2012), with a 50\,ns bunch crossing scheme
and a proton beam energy of 3.5\,(4)\,TeV. The higher instantaneous luminosity,
and only half of all bunches being filled, translates into a four times larger
occupancy per event as compared to the conditions that correspond to the design
parameters.

In Sec.~\ref{sec:hardware} the performance of the services is described
in terms of the gas quality, and the low and high voltage stability. In
Sec.~\ref{sec:commissioning} the performance of the electronics readout is
discussed, in particular the noise, amplifier threshold uniformity and
time-linearity.  The drift-time calibration and position alignment is shown in
Sec.~\ref{sec:calibration}.  The final detector performance in terms of
occupancy, single hit efficiency, resolution and radiation hardness is given in
Sec.~\ref{sec:performance}.

\section{Services performance}
\label{sec:hardware}

\subsection{Gas system}
The counting gas for the straw tube detectors of the OT 
was originally chosen as an admixture of Ar/CO$_2$/CF$_4$.
Studies on radiation resistance first suggested to operate without
CF$_4$~\cite{Bachmann:2004ir}, and subsequently with the addition
of O$_2$~\cite{Bachmann:2010zz}, leading to the final mixture
Ar/CO$_2$/O$_2$ (70/28.5/1.5).  
This choice is based on the requirement to achieve a reasonably fast charge
collection to cope with the maximum bunch crossing rate of 40\,MHz at the LHC, a
good spatial resolution and to maximize the lifetime of the detectors.

\begin{table}[!b]
\begin{minipage}{\linewidth}
\renewcommand{\footnoterule}{}
\renewcommand{\thefootnote}{\alph{footnote}}
\centering
\begin{tabular}[ht]{lcc}
  \hline
   & typical values & specifications \\
  \hline
  Gas flow                      & 800 - 850\,l/h   & $<1000$\,l/h           \\
  Overpressure in detector      & $1.6$\,mbar      & $< 5$\,mbar            \\
  Impurity (H$_2$O content)     & $<10$\,ppm       & $<50$\,ppm             \\
\hline  
\end{tabular}
\caption{\label{tab:gas}\small Main parameters of the OT gas system.}
\end{minipage}
\end{table}

The gas is supplied by a gas system~\cite{Barillere:2007zz} operated in an open
mode, without recycling of the gas. The gas system is a modular system, with the
mixing module on the surface and two distribution modules, a pump module, an
exhaust module and an analysis module in the underground area behind the
shielding wall to allow access during beam operation.  The gas is split between
two distribution modules, each supplying a detector half using 36 individual gas
lines. For each gas line the input flow can be adjusted and is measured
continuously, as well as the output flow. Each distribution module regulates the
pressure in the detector modules.  The analysis module allows to sample each of
the 36 lines individually at the detector inlet and outlet.  An oxygen sensor
and a humidity sensor are connected to the analysis rack. The measurement of one
gas line takes a few minutes such that each line is measured approximately once
every two hours.

The main operational parameters of the gas system are shown in
Table~\ref{tab:gas}.  The gas flow is kept low to prevent ageing effects
observed in laboratory measurements (see Sec.~\ref{sec:ageing}).  The detector
modules have been tested to be sufficiently gas-tight, on average below
$1.25\times 10^{-4}$~l/s (corresponding to 5\% gas loss every 2
hours)~\cite{Vankov:2012}, to prevent the accumulation of impurities from the
environment. The level of impurities is monitored by measuring the water content
in the counting gas, which is at a level below $10$\,ppm.

A system with pre-mixed bottles containing in total about 100~m$^3$ of
Ar/CO$_2$/O$_2$, is automatically activated in case of electrical power failures
of the main gas system, ensuring a uninterupted flow through the detector at all
times.  The gas mixture, and the level of impurities (H$_2$O) were stable during
the whole operation from 2010 to 2012.

\subsection{Gas monitoring}
\label{ss: gas_monitoring}

The gas quality for the OT is crucial, as it directly affects the detector gain
and stability, and potentially the hit efficiency and drift-time calibration.
Moreover, a wrong gas mixture can lead to accelerated radiation damage or
dangerously large currents.  The gas gain is determined with the help of two
custom built OT modules, of 1\,m length, which are irradiated by a $^{55}$Fe
source.

\begin{figure}[!b]                                                             
    \begin{picture}(250,150)(0,0)
    \put(100,0){\includegraphics[width=.48\textwidth]{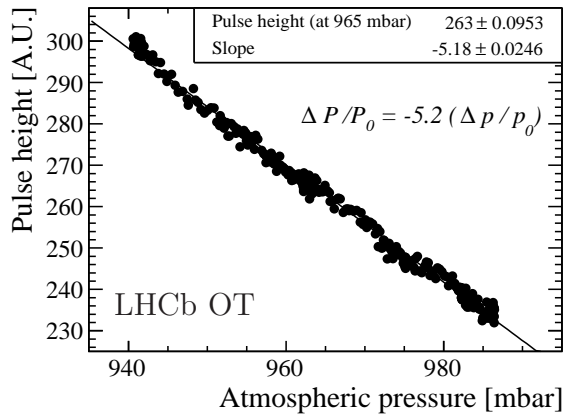}}
    \put(140,40){LHCb OT}
    \end{picture}
 \caption{\small
Pressure calibration curve of the $^{55}$Fe spectrum, obtained from the
dependence of the pulse height $P$ as a function of atmospheric pressure $p$.}
 \label{fg:Trabond_module}   
\end{figure}

One of the monitoring modules that was used in 2011 was constructed using a
particular glue, Trabond 2115, that does not provoke gain loss after long-term
irradiation with the $^{55}$Fe source.  The other module was built with the glue
used in mass production, Araldite AY103-1.  The modules were half-width modules
containing 32 straw tubes. The readout electronics consists of a high voltage
board carrying a number of single-channel charge pre-amplifiers.

The $^{55}$Mn K-$\alpha$ line of the $^{55}$Fe source has an energy of 5.9~keV
which is used as calibration reference.  Two $^{55}$Fe sources with low
intensity were used, resulting in a few events per second for both modules.  The
data acquisition system is based on a multi-functional readout box containing
two fast ADC inputs to which the amplified signals are fed. In addition the
atmospheric pressure is recorded. The $^{55}$Fe pulses are integrated over 15
minutes, and subsequently analyzed.  A double Gaussian distribution is fitted to
the $^{55}$Fe spectrum.  The peak position is then corrected for the atmospheric
pressure.

The pressure is measured inside a buffer volume at the chambers input. 
The pressure correction is determined from a linear fit to 
the  mean pulse-height as a function of the absolute atmospheric pressure, 
see Fig.~\ref{fg:Trabond_module}. The resulting stability of the gas gain
is within $\pm 2\%$ over 10 days. However, the gain loss due to ageing of
the monitoring modules was about 5\% over a period of about two months.
The monitoring modules were therefore replaced at the end of 2011 
by stainless steel modules, sealed with O-rings (instead of the standard
construction with Rohacell panels with carbon-fiber facing, glued together
to ensure the gas-tightness). The measurements of the gas gain
were stable throughout the entire running period of 2012.

\subsection{Low voltage}
The function of the low voltage (LV) distribution system is to provide the bias
voltages to the front-end electronics.  Each FE box (the GOL/AUX board) hosts
three radiation-hard linear voltage regulators ($+2.5\,\mathrm{V}$,
$+3\,\mathrm{V}$ and $-3\,\mathrm{V}$) biased by two main lines,
$+6\,\mathrm{V}$ and $-6\,\mathrm{V}$.  Two distribution boxes per C-frame split
the $+6\,\mathrm{V}$ and $-6\,\mathrm{V}$ supply lines to the 18 FE boxes at the
top and the 18 FE boxes at the bottom; all supply lines to the FE boxes are
individually provided with slow fuses ($4\,\mathrm{A}$ for $+6\,\mathrm{V}$ and
$2\,\mathrm{A}$ for $-6\,\mathrm{V}$) and LED's showing their status.

The low-voltage distribution systems worked reliably throughout the 2010 to 2012
data taking periods.  In a few cases a single fuse of a FE box broke and was
replaced in short accesses to the LHCb cavern.

\subsection{High voltage}
The anode wires are supplied with +1550~V during operation, which corresponds to
a gas gain of about $5\times 10^4$~\cite{VanApeldoorn:2004jn}.  Each FE box has
four independent high voltage (HV) connections, one for each 32-channel HV
board.  Two mainframes~\footnote{CAEN SY1527LC \textregistered.}, each equipped
with four 28-channels supply boards~\footnote{A1833B PLC \textregistered.}, are
used as HV supply.  Using an 8-to-1 distribution scheme a total of 1680 HV
connections of the detector are mapped on 210 CAEN HV channels.  The
distribution is realized using a patch panel which offers the possibility to
disconnect individual HV boards by means of an HV jumper. Both components, the
HV supply as well as the patch panel, are located in the counting house. Access
to the HV system during data taking is therefore possible.

The typical current drawn by a single HV channel (supplying 256 detector
channels) depends on the location in the detector and varies between 20 and
150\,$\mu$A.  The short-circuit trip value per HV channel was set to 200\,$\mu$A
with the exception of one channel were the current shows an unstable behaviour,
and where the trip value was increased to 500\,$\mu$A.
The power supply can deliver a maximum current of 3~mA for a single HV channel.

In the 2011 and 2012 running periods there were 8 single detector channels
(wires) that showed a short-circuit, either due to mechanical damage, or due to
a broken wire.  During technical stops these single channels were disconnected,
to allow the remaining 31 detector channels on the same HV board to be
supplied with high voltage.

\section{Commissioning and monitoring}
\label{sec:commissioning}
Quality assurance tests of the detector modules, the FE-boxes and C-frame
services were performed prior to installation. Faulty components were repaired
whenever possible.  During commissioning and operating phases of the LHCb
detector, the stability and the quality of the OT \FE-electronics performances
was monitored.  Upon a special calibration trigger, sent by the readout
supervisor, the \FE-electronics generates a test-pulse injected via the \ASDBLR
test input~\cite{Dressnandt:2000fe,Gromov:2000}.  Test-pulse combinations can be
generated, and is implemented such that only even or only odd numbered channels,
or all channels simultaneously, are injected with charge.

\subsection{Quality assurance of detector modules and C-frame services}
The quality of the detector modules was assured by measuring the wire tension,
pitch, and leakage current (in air) prior to the module sealing.  Subsequently,
the gas tightness of the detector module was measured.  Finally, the
functionality of each wire was validated in the laboratory immediately after
production, by measuring the response to radioactive sources ($^{55}$Fe or
$^{90}$Sr)~\cite{VanApeldoorn:2004jn}.

Before and after shipment of the C-frames from Nikhef to CERN, the equipment of
the services were checked, namely the gas tightness of the gas supply lines, the
dark currents on the high-voltage cables, the voltage drop on the low-voltage
supply cables and the power attenuation of the optical
fibers~\cite{Simioni:2010zz,Jansen:2011}.

Following the installation of the modules on the C-frames in the LHCb
cavern, the gas tightness of each module was confirmed. The response to
$^{55}$Fe of approximately half of the straws was measured
again, resulting in 12 noisy channels, 7 dead channels, and 4
straws with a smaller gas flow~\cite{Jansen:2011}.

All FE-boxes were measured on a dedicated test-stand in the laboratory, and the
faulty components were replaced. The tests performed on the test-stand are
identical to the tests that are performed regularly during running periods.  
Three sequences of test runs are provided:
\begin{itemize}
\item runs with a random trigger at varying \ASDBLR 
threshold settings to measure the noise rate; 
\item runs with test-pulse injected at the ASDBLR test
input, at varying threshold settings to check the full readout chain for
threshold uniformity and cross-talk; 
\item runs with test-pulse (at
fixed threshold) at increasing test-pulse delay settings, to determine the
time-linearity.
\end{itemize}

\subsection{Noise}
The noise scan analysis aims at identifying channels that have an ``abnormal''
level of noise, which may be due to dark pulses from the detector, bad
\FE-electronics shielding, or bad grounding.  In each channel the fraction of
hits is determined for increasing values of the amplifier threshold, triggered
randomly. The nominal value of the amplifier threshold is 800\,mV, which
corresponds to an input charge of about 4\,fC.

 \begin{figure}[!b]
   \begin{picture}(250,150)(0,0)
     \put(0,0){\includegraphics[width=.5\textwidth]{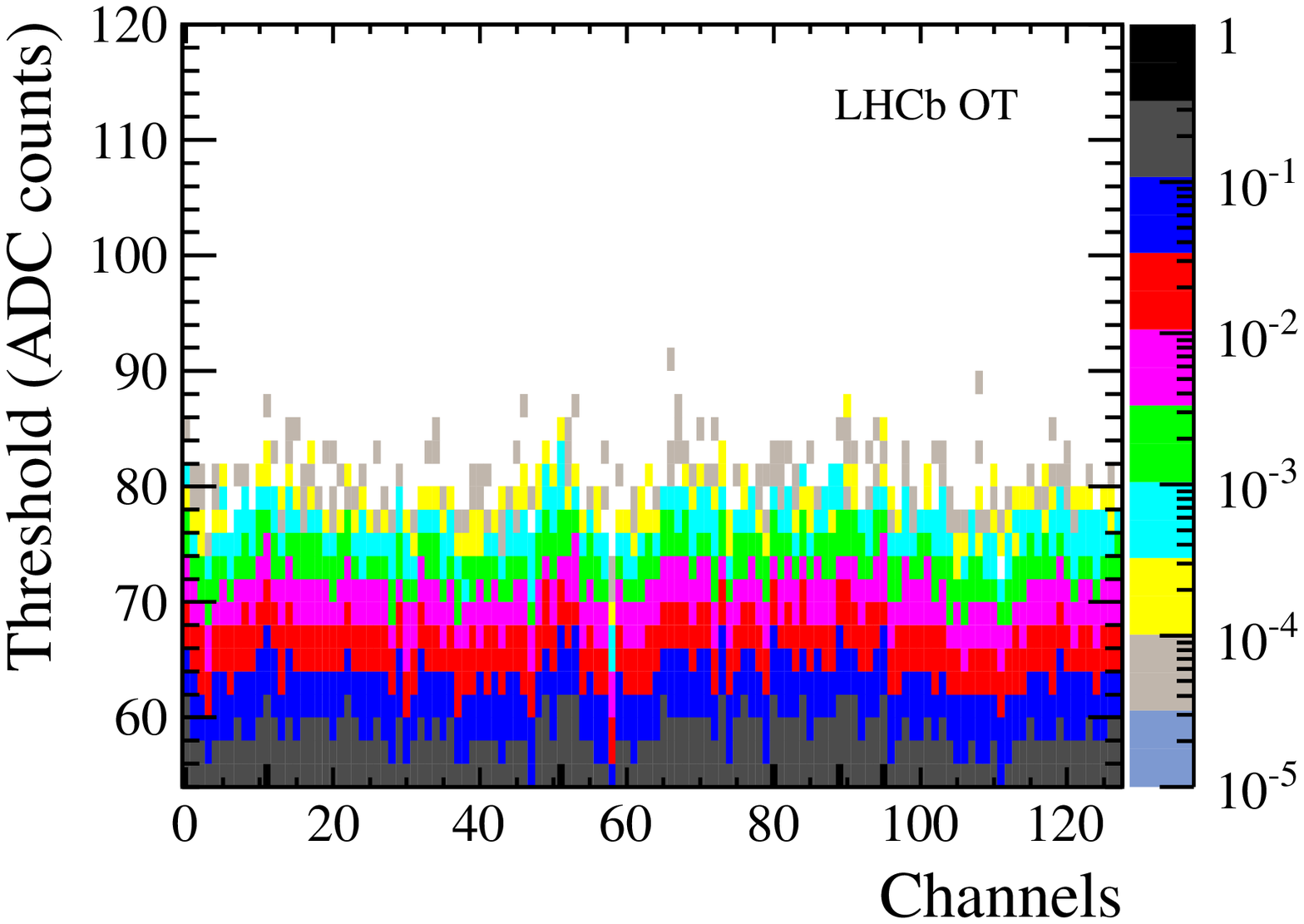}}
     \put(230,0){\includegraphics[width=.5\textwidth]{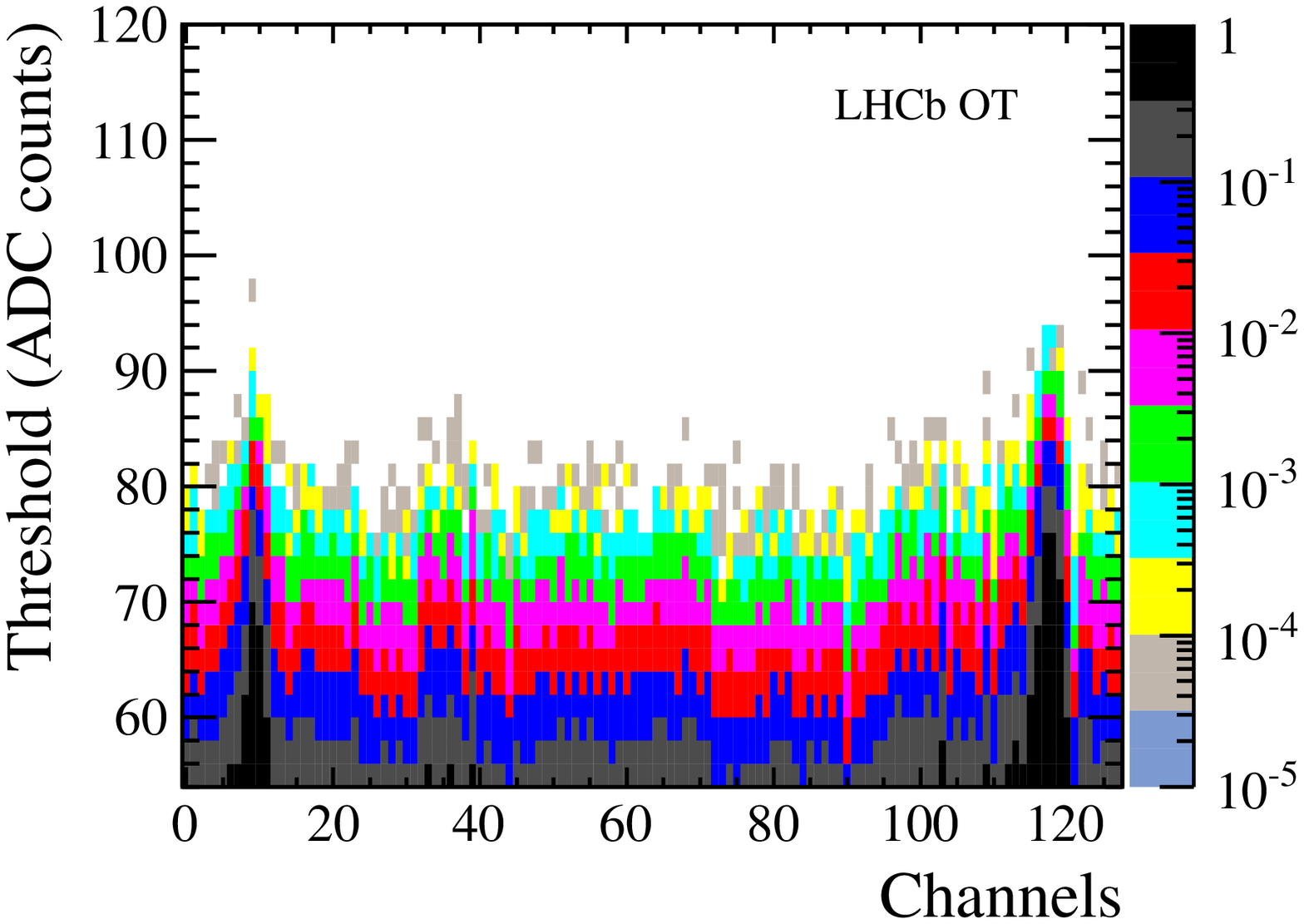}}
    \put(0,0){(a)}
    \put(230,0){(b)}
   \end{picture}
 \caption{\small The 2d-hitmap histogram showing the noise occupancy, 
   for each channel, and varying amplifier
   threshold (1~ADC count $\approx$ 10~mV)~\cite{Storaci:2012} for
   (a) a typical FE-box with  good channels and 
   (b) a FE-box with two groups of noisy channels.}
\label{fig:noisyEx}
\end{figure}

The typical noise occupancy for 128 channels in one FE box is shown in
Fig.~\ref{fig:noisyEx}(a) for increasing amplifier threshold, where the
occupancy is defined as the ratio of the number of registered hits in that
channel over the total number of triggered events.  A noise occupancy at the
level of $10^{-4}$ is observed at nominal threshold, as expected from beam
tests~\cite{vanApeldoorn:2005ss}.  These results are representative for about
98\% of all FE boxes in the detector.

An example of a FE box with a few noisy channels is shown in
Fig.~\ref{fig:noisyEx}(b), where two groups of about 5 noisy channels are
identified. About 2\% of the FE boxes exhibited this noise pattern during the
2011 running period.
At the nominal threshold of 800~mV (4~fC) only 0.2\% of the channels
exhibited a noise occupancy larger than 0.1\%.
During the 2011/2012 winter shutdown, this noise pattern was understood and 
identified to be caused by imperfect grounding, and was subsequently solved.

\subsection{Threshold scans}
\label{ss:threshold_scan}
The threshold scan records hits at fixed input charge 
(given by either a low or high test-pulse 
of 4 and 12~fC, respectively) and is aimed at monitoring the 
gain of the \FE-electronics preamplifier, in order to locate dead channels,
determine gain deteriorating effects and measure cross-talk.

\begin{figure}[!t] 
   \begin{picture}(250,150)(0,0)
     \put(0,0){ \includegraphics[width=.5\textwidth]{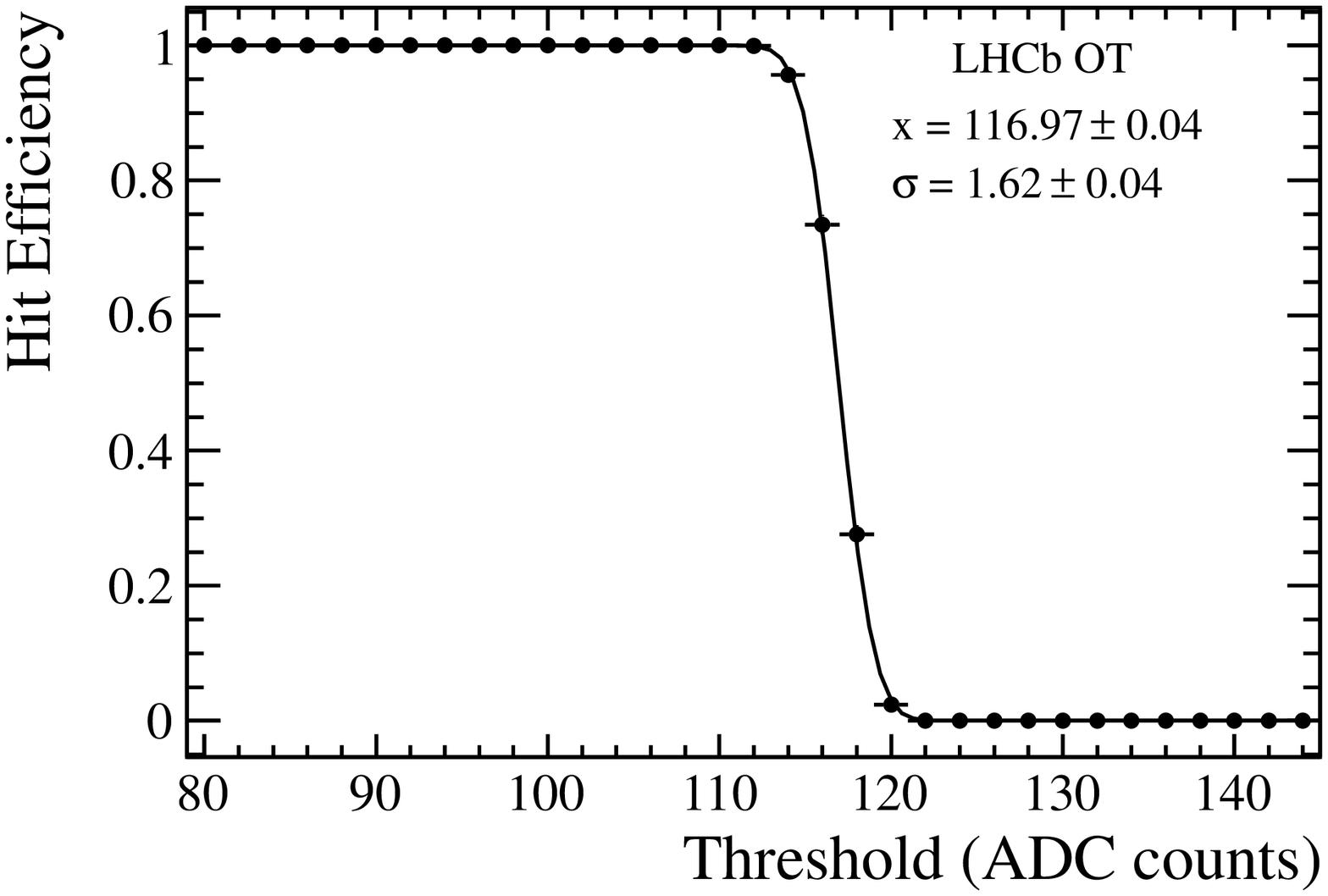}}
     \put(230,0){\includegraphics[width=.5\textwidth]{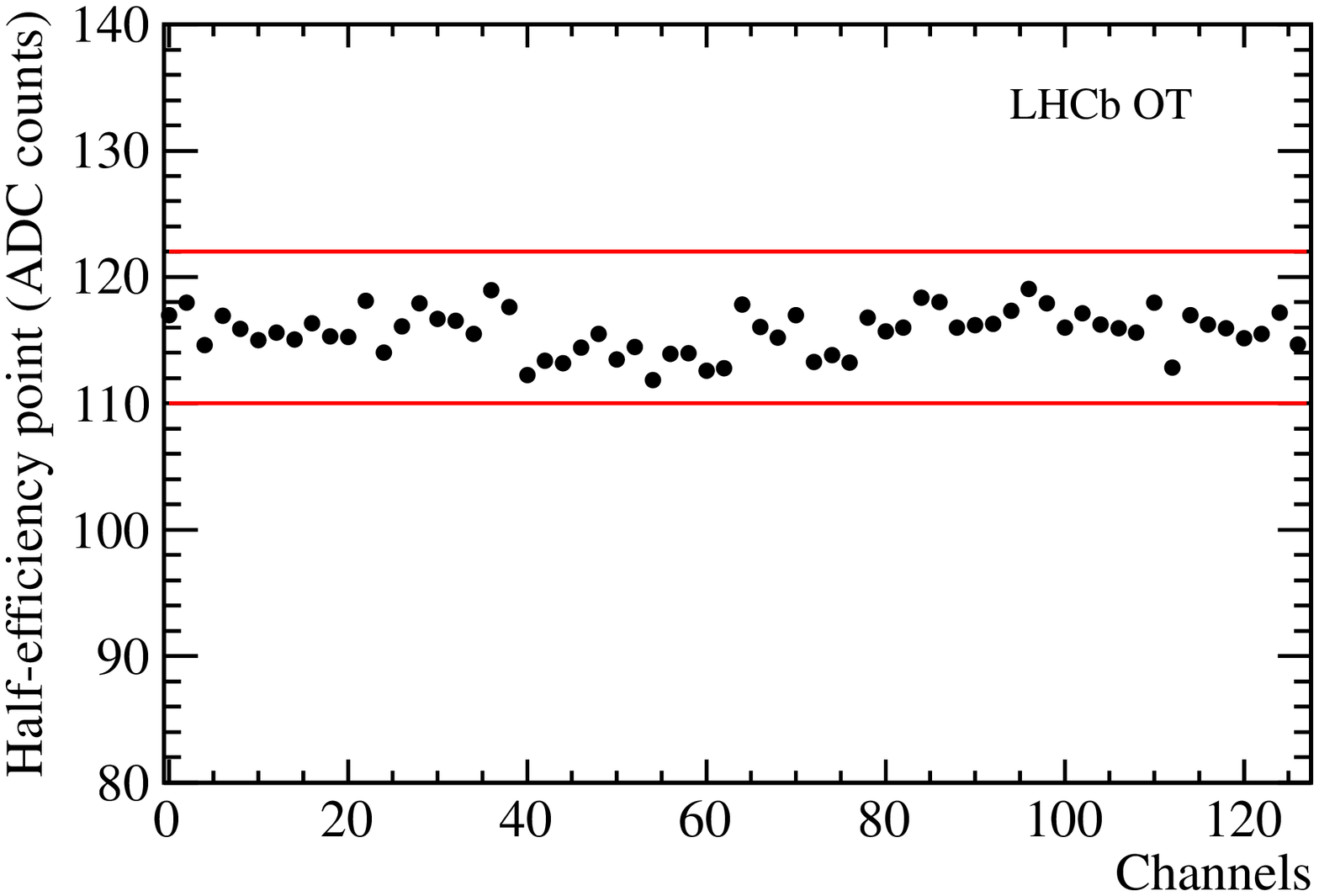}}
     \put(0,0){(a)}
     \put(230,0){(b)}
   \end{picture}         
 \caption{\small 
 (a) Example of hit-efficiency as function of threshold for a fixed
 input charge (``high test-pulse'')~\cite{Storaci:2012}.
 (b) Stability of the half-efficiency point for channels in one FE-box 
 (1~ADC count $\approx$ 10~mV).}
 \label{fg:thrScan}                                                            
\end{figure}

The \ASDBLR chip selection prior to assembly of the \FE-box components 
guarantees a good uniformity of the discriminators, such that a common
threshold can be applied for the entire readout without loss of
efficiency or increased noise levels~\cite{Simioni:2010zz}.
An error function produced by the convolution of a step function (ideal
condition in absence of noise) with Gaussian noise, is used to describe
the hit-efficiency as a function of the threshold value. 
The stability of the half-efficiency 
point for all the channels was studied and the relative variation between 
channels is expected to be less than $\pm60$~mV~\cite{Simioni:2010zz}.

An example of the fit to the hit efficiency as a function of amplifier threshold
is shown for one channel in Fig.~\ref{fg:thrScan}(a), and the half-efficiency
point for 128 channels in one FE box is shown in Fig.~\ref{fg:thrScan}(b).
Since the start of the data taking period in May 2010 the fraction of fully
active channels has been 99.5\% or more, see Sec.~\ref{sec:badch}.

\subsection{Delay scans}
\label{ss:delay_scan}
\begin{figure}[!t]
  \begin{picture}(250,150)(0,0)
    \put(0,0){\includegraphics[width=.5\textwidth]{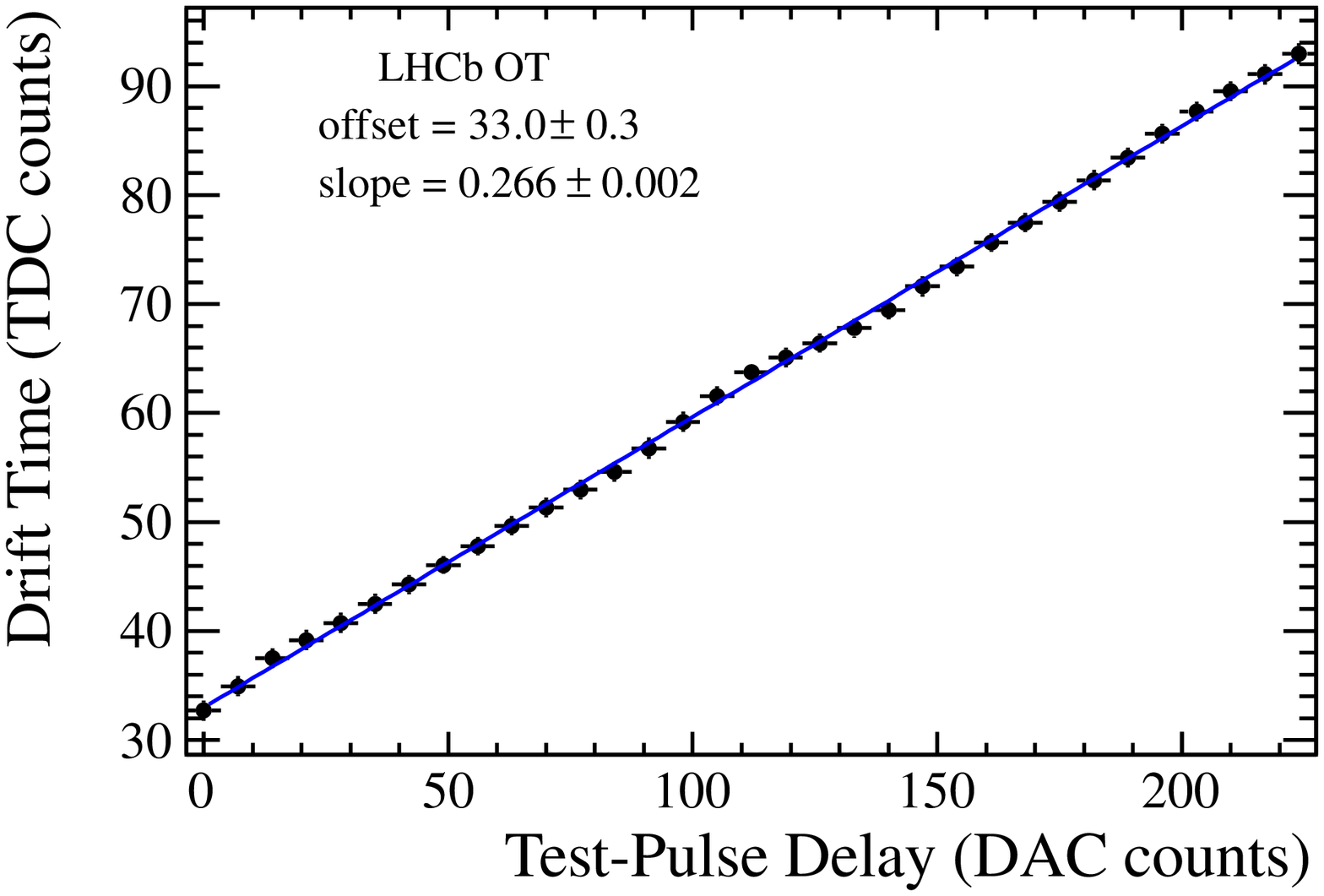}}
    \put(230,0){\includegraphics[width=.5\textwidth]{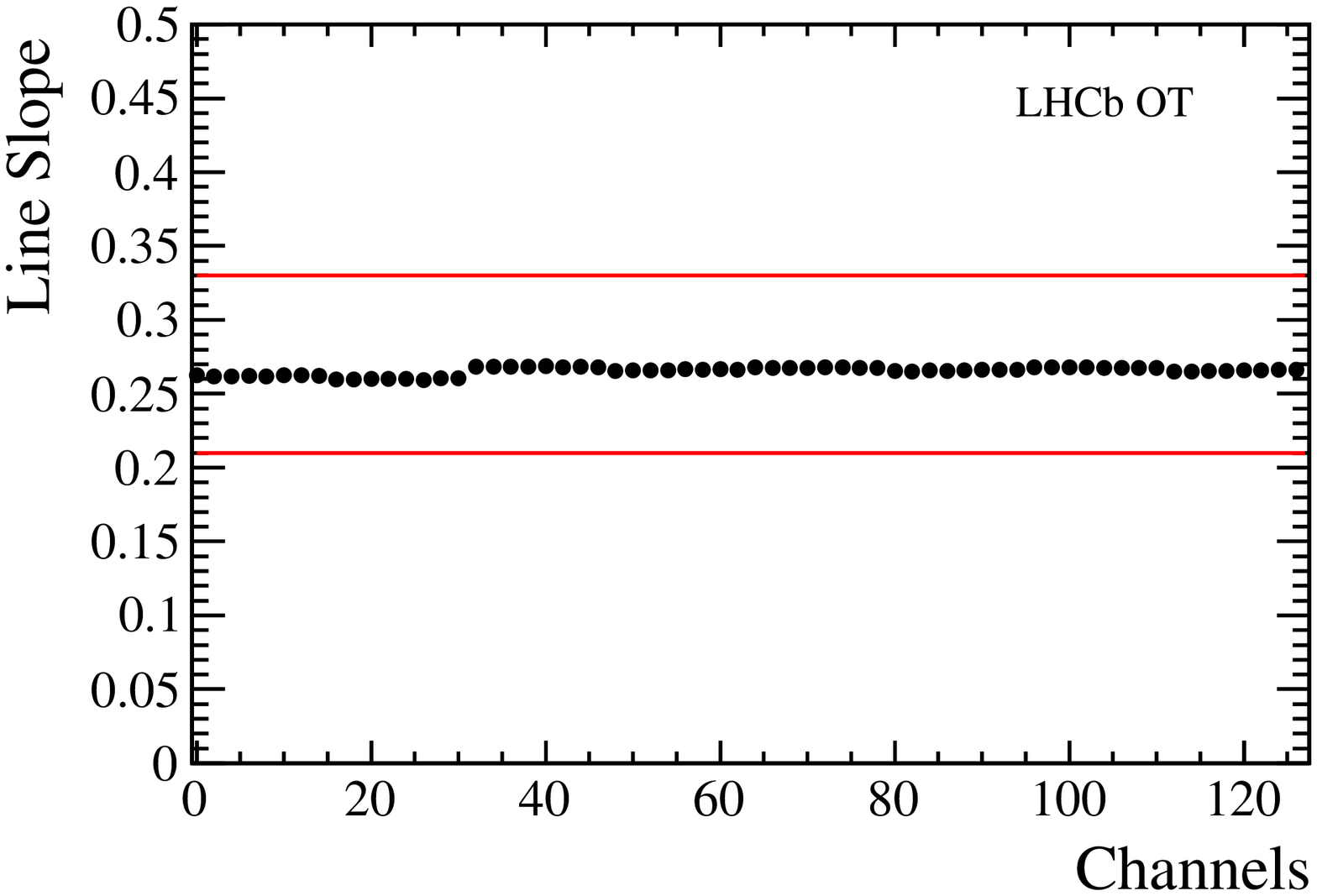}}
    \put(0,0){(a)}
    \put(230,0){(b)}
  \end{picture}         
 \caption{\small 
 (a) Example of a linear fit of the measured drift-time as a function of the
 test-pulse delay~\cite{Storaci:2012}.  The slope corresponds to unity, if both
 axis are converted to ns (1~DAC count $\approx$ 0.1~ns, while 1~TDC count
 $\approx$ 0.39~ns).  
 (b) The slope from the linear fit of the timing measurement
 for all 128 channels in one FE-box.}
\label{fg:delayScan}
\end{figure}

The delay scan analysis aims at detecting defects in the timing of the OT
channels, such as time offsets or non-linearities.  An example of the time
measurement as a function of the test-pulse delay is shown in
Fig.~\ref{fg:delayScan}(a), where each measurement corresponds to the average of
10,000 time measurements at a given test-pulse delay.  The corresponding slope
for all 128 channels of one FE box is shown in Fig.~\ref{fg:delayScan}(b).  No
anomalous behaviour in the time measurement has been observed.  Approximately
96\% of all channels have a slope between 0.983 and 1.024 times the average
value.  Most of the remaining 4\% of the channels suffer from insufficient
test-pulse stability rather than from pre-amplifier or TDC shortcomings.

\newcommand{\tclock}{\ensuremath{t_{clock}}\xspace}
\newcommand{\tgate}{\ensuremath{t_{gate}}\xspace}
\newcommand{\tzero}{\ensuremath{t_{0}}\xspace}

\section{Calibration}
\label{sec:calibration}
The position of the hits in the OT is determined by measuring the drift-time to
the wire of the ionisation clusters created in the gas volume. The drift-time
measurement can in principle be affected by variations in the time offset in the
FE electronics, and is regularly monitored.  The spatial position of the OT
detector also affects the hit position, and the correct positioning of the
detector modules is ensured by periodic alignment campaigns.

\subsection{Distance drift-time relation}
\label{sec:tr}

\begin{figure}[!t]
    \begin{picture}(250,250)(0,0)
     \put(70,5){\includegraphics[bb=0 0 544 309,scale=0.7]{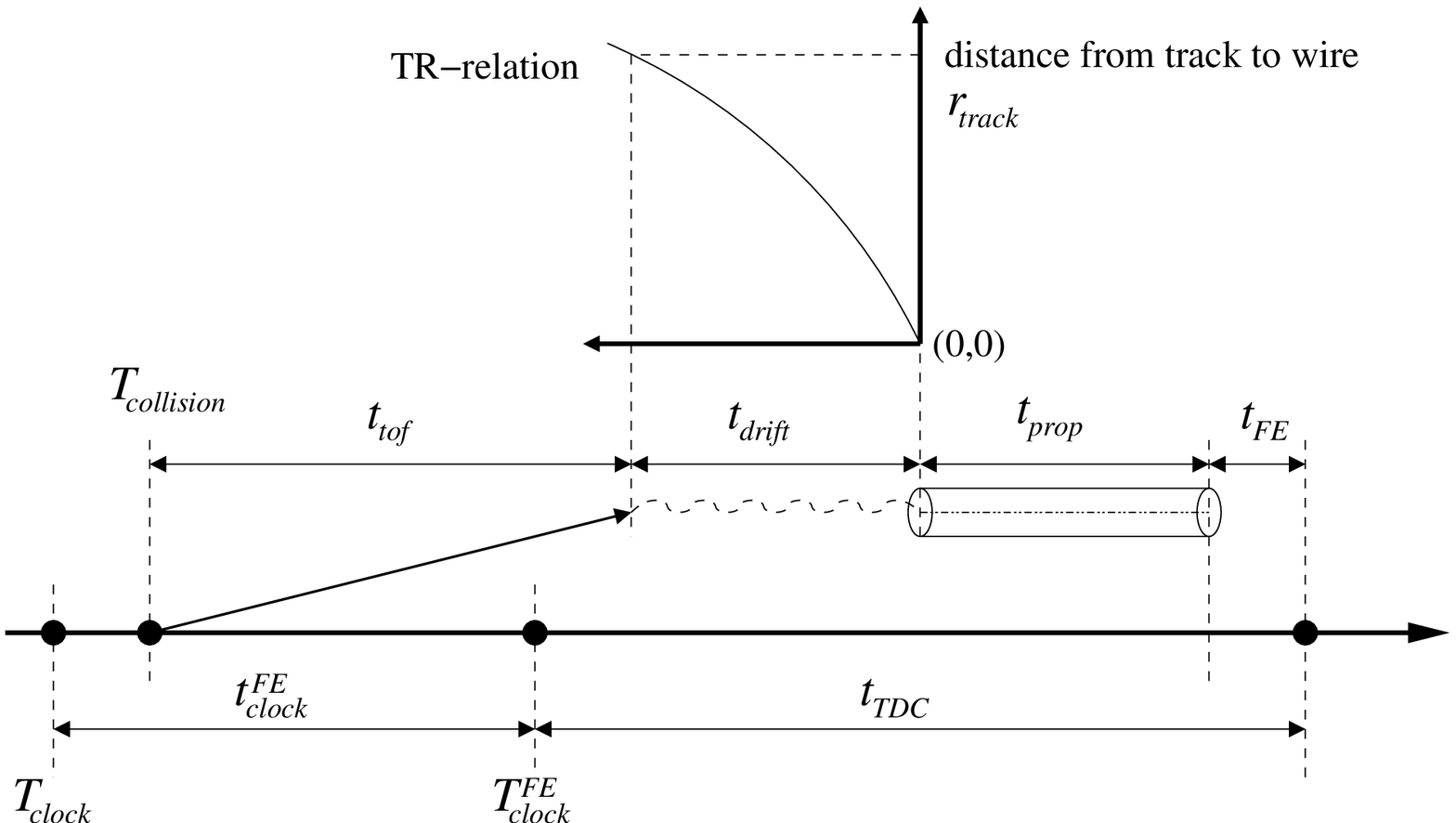}}
     \put(-10,125){\includegraphics[bb=0 0 284 230,scale=0.5]{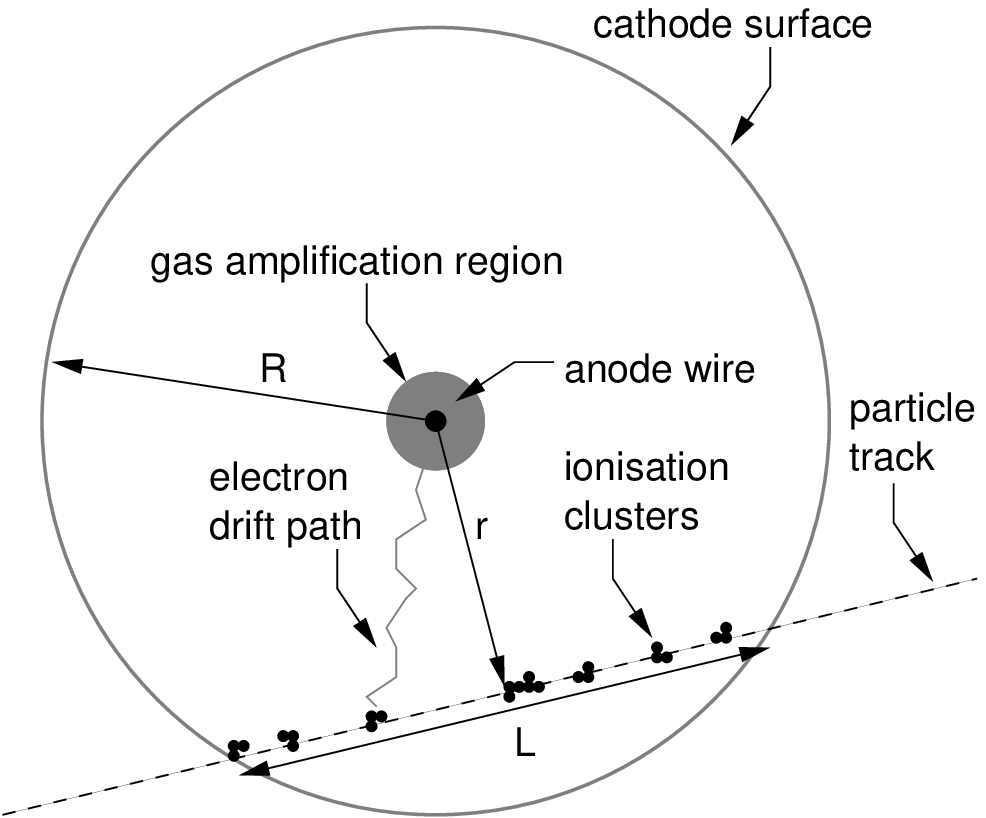}}
     \put(0,100){(b)}
     \put(0,10){(a)}
     \end{picture}	
\caption[Sketch]{\small
  \label{fig:tdc:sketch}
    (a) Sketch of the various contributions to the measured TDC 
    time~\cite{Kozlinskiy:2013}, as explained in the text.
    (b) Picture of a charged particle that traverses a straw.}
\end{figure}

The OT detector measures the arrival time of the ASDBLR signals with respect to
the LHC clock, $T_\mathrm{clock}$, and is referred to as the TDC time,
$t_\mathrm{TDC}$. This time is converted to position information to reconstruct
the trajectory of the traversing charged particle, by means of the
drift-time--distance relation, or TR-relation.  The arrival time of the signal
corresponds to the time of the $pp$ collision, $T_\mathrm{collision}$, increased
by the time-of-flight of the particle, $t_\mathrm{tof}$, the drift-time
$t_\mathrm{drift}$ of the electrons in the straw, the propagation time of the
signal along the wire to the readout electronics, $t_\mathrm{prop}$, and the
delay induced by the FE electronics, $t_\mathrm{FE}$. The various contributions
to the TDC time are schematically shown in Fig.~\ref{fig:tdc:sketch}, and can 
be expressed as

\begin{equation}
t_\mathrm{TDC}  = 
(T_\mathrm{collision} - T^\mathrm{FE}_\mathrm{clock}) + 
t_\mathrm{tof} + t_\mathrm{drift} + t_\mathrm{prop} + t_\mathrm{FE}. 
\end{equation}
The phase of the clock at the TDC input, $T^\mathrm{FE}_\mathrm{clock}$,
can be adjusted with a shift $t^\mathrm{FE}_\mathrm{clock}$.
The expression for $t_\mathrm{TDC}$ can be rewritten as 

\begin{equation}
t_\mathrm{TDC}  = 
(T_\mathrm{collision} - T_\mathrm{clock}) + t_0
+ t_\mathrm{tof} + t_\mathrm{drift} + t_\mathrm{prop},
\end{equation}
where  $t_0 = t_\mathrm{FE} - t^\mathrm{FE}_\mathrm{clock}$.
Variations in $t_0$ are discussed in the next section.  The difference
$t_\mathrm{clock} = T_\mathrm{collision} - T_\mathrm{clock}$ accounts for variations of the phase
of the LHC clock received at the LHCb experiment control and is kept below
0.5~ns.

The TR-relation is the relation between the measured drift-time and the closest
distance from the particle trajectory to the wire.  The TR-relation is
calibrated on data by fitting the distribution of drift-time as a function of
the reconstructed distance of closest approach between the track and the wire,
as shown in Fig.~\ref{fig:clbr:tr}(a).  At the first iteration the TR-relation
obtained from beam tests was used.  The line shows the currently used
TR-relation~\cite{Kozlinskiy:2013}, which has the following parameterization:

\begin{figure}[!t]
    \begin{picture}(250,420)(0,0)
     \put(60, 190){\includegraphics[bb=0 0 530 420,scale=0.5]{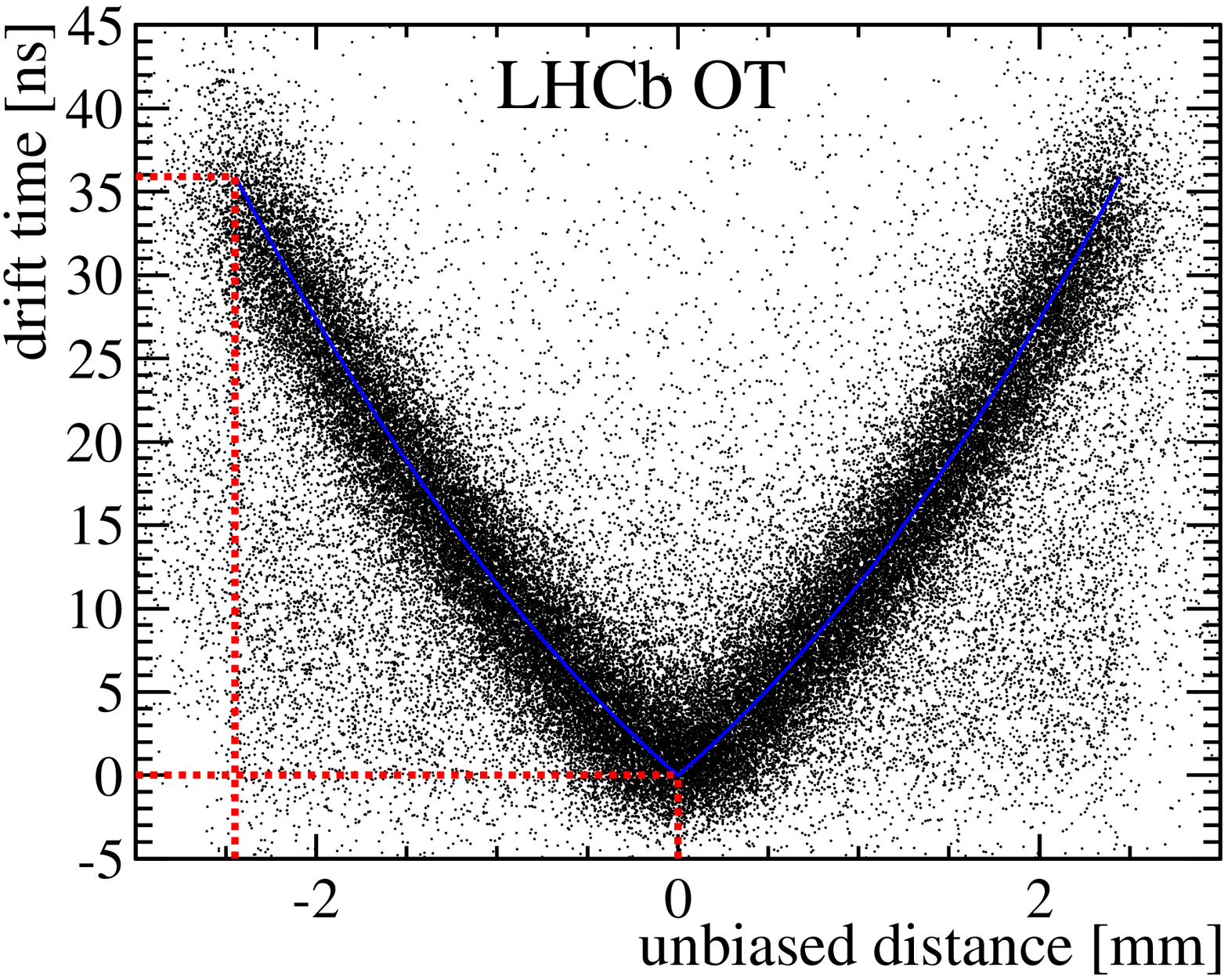}}
     \put(-10,  5){\includegraphics[bb=0 0 565 423,scale=0.4]{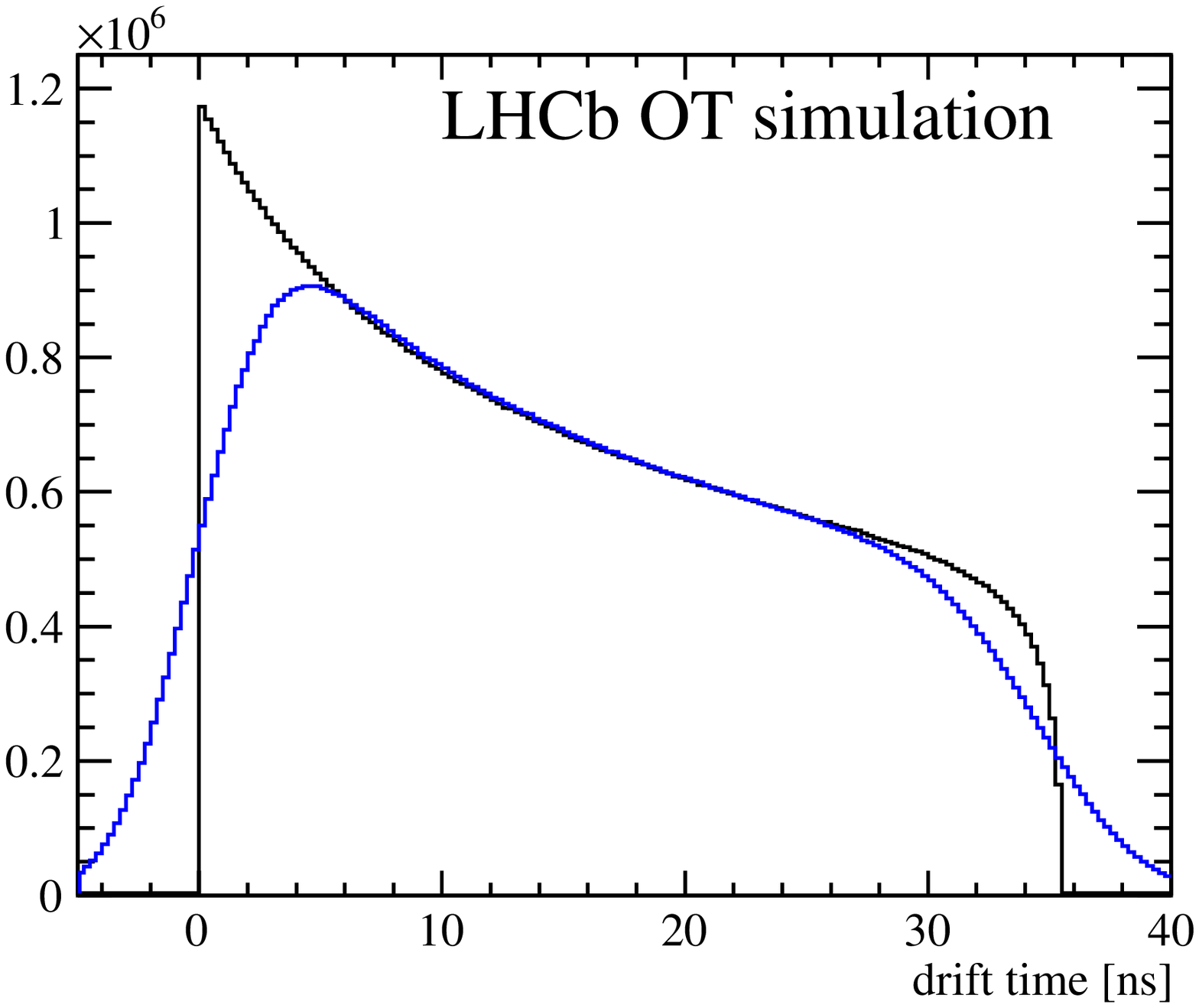}}
     \put(220,  5){\includegraphics[bb=0 0 565 423,scale=0.4]{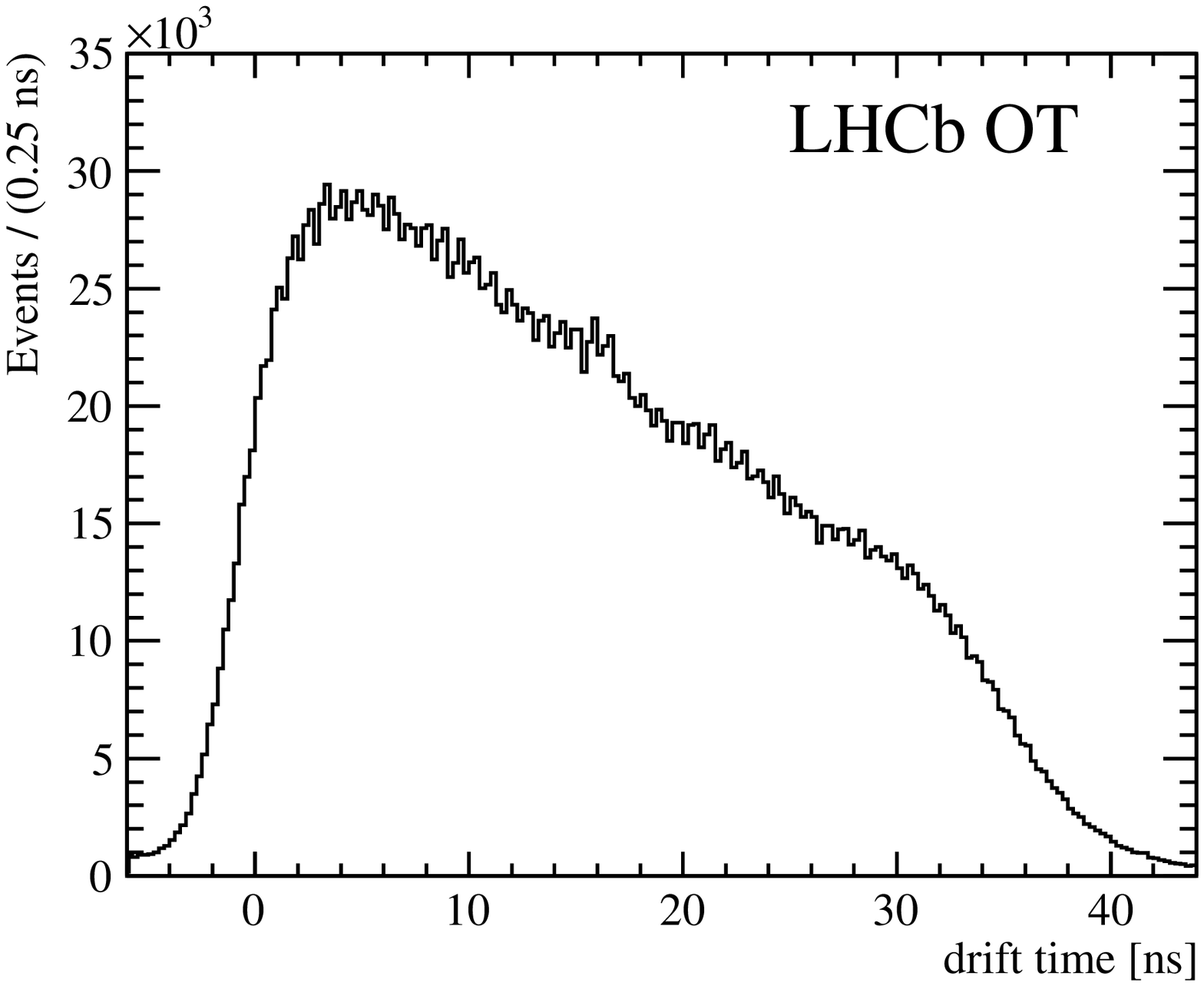}}
     \put(70, 185){(a)}
     \put(10,0){(b)}
     \put(230,0){(c)}
     \end{picture}	
\caption[TR-relation distribution]{\small
  \label{fig:clbr:tr}
  The (a) TR-relation distribution follows the shape of a second order
  polynomial distribution, which leads to a
  (b) falling drift-time spectrum (black), which,
  smeared with the time resolution (blue), leads to the shape of the  
  (c) measured drift-time distribution.}
\end{figure}

\begin{equation}
t_\mathrm{drift}(r) 
= 20.5\,\mathrm{ns} \cdot \frac{|r|}{R} 
+ 14.85\,\mathrm{ns} \cdot \frac{r^2}{R^2} \text{,}
\end{equation}
\noindent where $r$ is the closest distance between the track and the wire and 
$R = 2.45$\,mm is the inner radius of the straw.  This TR-relation is compatible
with the one obtained from the beam test of 2005~\cite{vanApeldoorn:2005ss},
$t(r) = 20.1\,\mathrm{ns} \cdot \frac{|r|}{R} + 14.4\,\mathrm{ns} \cdot
\frac{r^2}{R^2}$.  The maximum drift-time extracted from the parameterization of
the TR-relation is 35\,ns. Due to the average drift-time resolution of 3\,ns,
and due to the variation in time-of-flight of the traversing particles, the
drift-time distribution broadens, as illustrated in Fig.~\ref{fig:clbr:tr}(b).  
The measured drift-time spectrum after $t_0$
calibration is shown in Fig.~\ref{fig:clbr:tr}(c), and the start of
the drift-time spectrum is thus set to 0\,ns by construction.
During operation, the start of the 75\,ns wide readout gate was set to
approximately $-9$\,ns, to ensure that also the earliest hits are recorded.  The
varying number of entries in the subsequent time bins is a characteristic of the
OTIS TDC chip known as the differential non-linearity (caused by variations of the 
digital delay bin sizes) and does not significantly affect the drift-time 
resolution~\cite{Jansen:2011}.

\subsection{\texorpdfstring{\tzero} ~stability}

\begin{figure}[!b]
    \begin{picture}(250,170)(0,0)
     \put(10,0){\includegraphics[bb=0 0 566 213,scale=0.75]{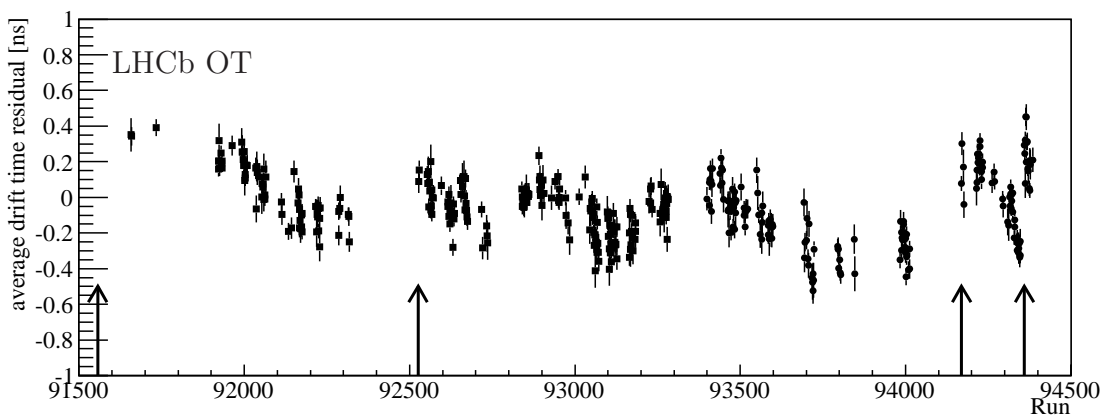}}
     \put(50,130){LHCb OT}
      \end{picture}	
\caption[\tzero stability versus run number]{\small
  \label{fig:clbr:stability:run}
  \tzero stability versus run number.
  Every point corresponds to one run that typically lasts one hour. 
  The arrows indicate the adjustment of the \tclock time. 
}
\end{figure}
\begin{figure}[!t]
    \begin{picture}(250,210)(0,0)
     \put(60,0){\includegraphics[bb=0 0 565 382,scale=0.5]{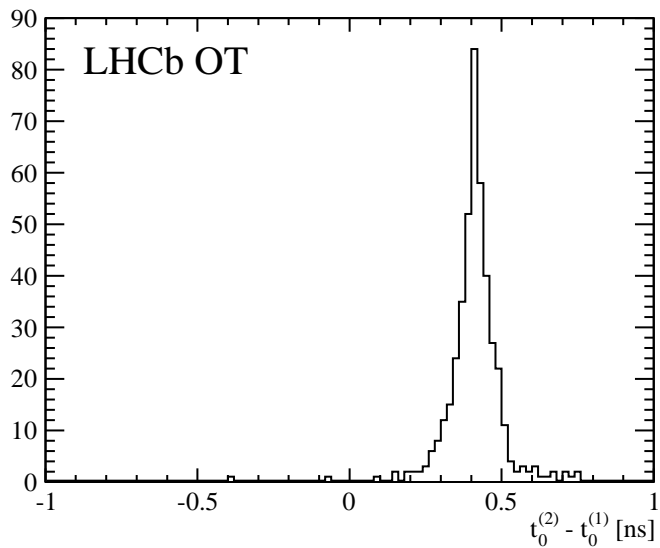}}
     \end{picture}	
\caption[\tzero stability versus run number]{\small
  \label{fig:clbr:stability:distr}
  Distribution of differences between \tzero constants per FE box, for two
  different calibrations. The mean shift originates from a change of the overall
  \tclock time, whereas the spread shows the stability of the delay 
  $t_\mathrm{FE}$ induced by the FE electronics.
}
\end{figure}

The $t_\mathrm{FE}$ values need to be stable to a level better than the time
resolution.  There are two factors that contribute to the stability of
$t_0$, usually referred to as the \tzero constants: one is the drift
of the global LHCb clock and the second is the drift of FE electronic
delays. The first can be extracted from the average over the whole OT of the
drift-time residual distribution calculated for every run separately. The second
can be estimated from the difference of \tzero values for two different
calibrations, for each FE-box.

Figure \ref{fig:clbr:stability:run} shows the variation of the LHCb clock as a
function of the run number, for the data taking period between May and July
2011.  The global LHCb clock is adjusted if it changes by more than 0.5~ns.  As
a result, the average value of the drift-time residual stays within the range of
$\pm0.5$\,ns.

Figure \ref{fig:clbr:stability:distr} shows the difference of the
\tzero values per FE-box, 
for two different calibrations performed on runs 89350 and 91933, respectively.
These runs correspond to the beginning of two data taking periods in May and
July 2011. For most FE boxes the spread of the \tzero constants is smaller than
0.1\,ns.  The overall shift of 0.4\,ns is due to the change of the global LHCb
clock.  The variation of $t_0$ is well below the time resolution of
3~ns and does therefore not contribute significantly to the detector resolution.

\subsection{Geometrical survey}
The correct spatial positioning of the OT modules is ensured in three steps.
First, the design and construction of the OT detector guarantees a mechanical
stability of 100 (500)\,$\mu$m in the $x$($z$) direction.  Secondly, an optical
survey determined the position of all modules after installation.  Finally, the
use of reconstructed tracks allows to measure the position of the detector to
the highest accuracy.

By construction the anode wire is centered within 50\,$\mu$m with respect to the
straw tube.  The detector modules are fixed with dowel-pins to the C-frames at
the top and the bottom, with tolerances below 50\,$\mu$m. The modules are not
fixed at the center, making larger variations possible (see
Sec.~\ref{sec:align}).  Finally, the C-frames are mounted on rails, which fixes
the $z$-coordinate at the top and at the bottom.

First, the survey confirmed that the rails were straight within the few
millimeters tolerance.  Then, after installation, the position of the four
corners of the C-frames were adjusted until all measured points on the
dowel-pins at the top and bottom of the modules, and on the surface at the
center of the modules, were within $\pm 1$\,mm of their nominal position.  The
final survey coordinates provided the corrections to the nominal coordinates of
the C-frames and modules~\cite{Amoraal:2011}.  The C-frames can be opened for
maintenance, and the reproducibility of the C-frame positioning in the
$x$-coordinate was checked to be better than the 200\,$\mu$m precision of the
optical survey.  The shape of the modules in the $x$-coordinate is finally
determined using reconstructed tracks, see Sec.~\ref{sec:align}.

\subsection{Optical alignment with the Rasnik system}
The stability of the C-frame relative position during data taking is monitored
by means of the Rasnik system~\cite{Dekker:1993qq,rasnik:2001}.  The Rasnik
system consists of a CCD camera that detects a detailed pattern.  The pattern,
or ``mask'', is mounted on the C-frame and a movement is detected by the CCD
camera as a change of the pattern position.  All four corners of the 12 C-frames
are equipped with a Rasnik system.  Together with two additional Rasnik lines to
monitor movements of the suspension structure, this leads to a total of 50
Rasnik lines. Due to mechanical conflicts in the installation, only about 2/3 of
the lines are used.  The intrinsic resolution of the system perpendicular
(parallel) to the beam axis is better than 10 (150)\,$\mu$m.  The Rasnik
measurements showed that the position of the C-frames is unchanged after opening
and closing within $\pm 10$\,$\mu$m, and unchanged within $\pm 20$\,$\mu$m for
data taking periods with opposite polarity of the LHCb dipole magnet.

\subsection{Software alignment}

\begin{figure}[!b]
   \begin{picture}(250,170)(0,0)
     \put(10, 0){\includegraphics[width=0.41\textwidth]{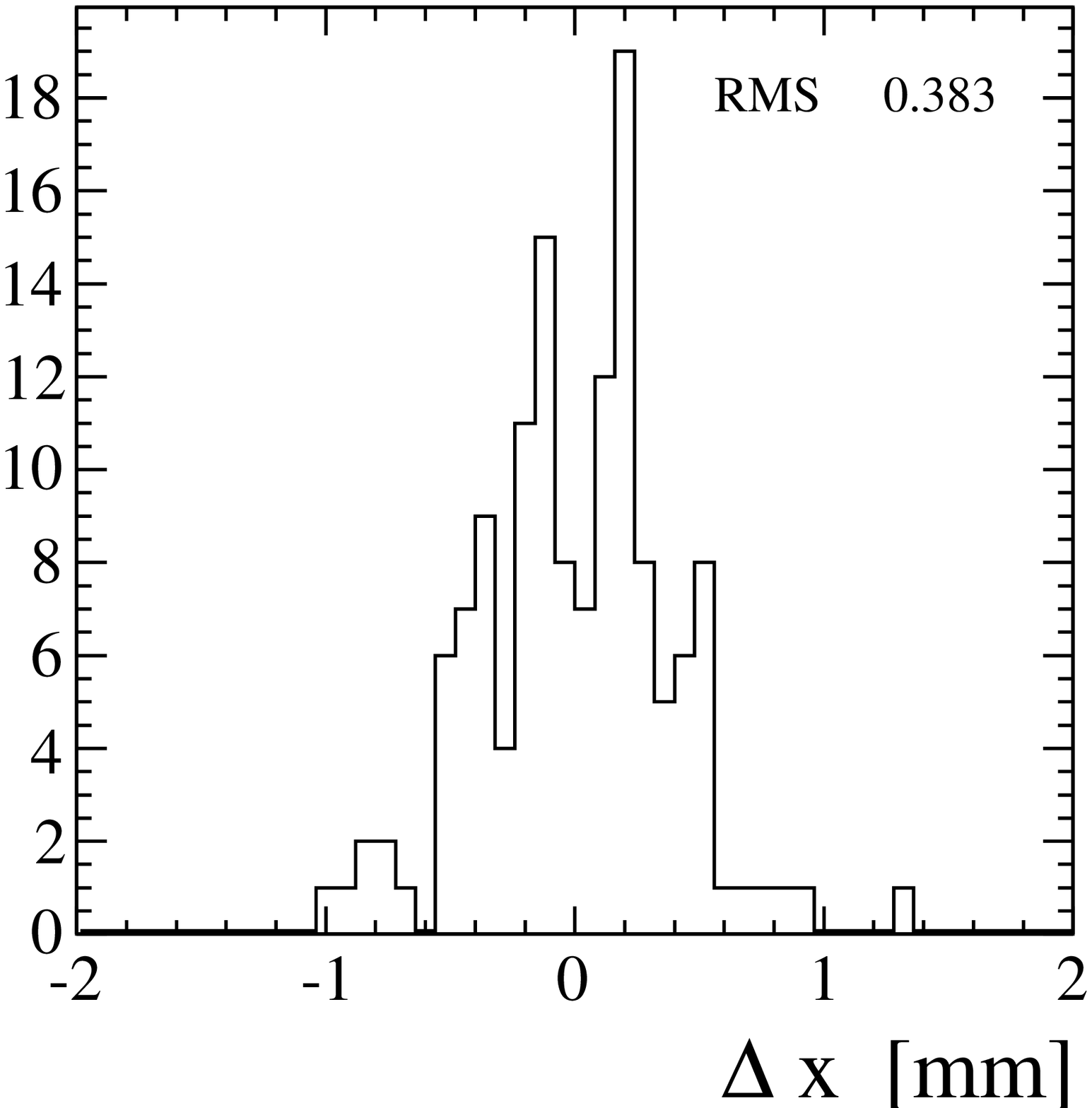}}
     \put(240,0){\includegraphics[width=0.41\textwidth]{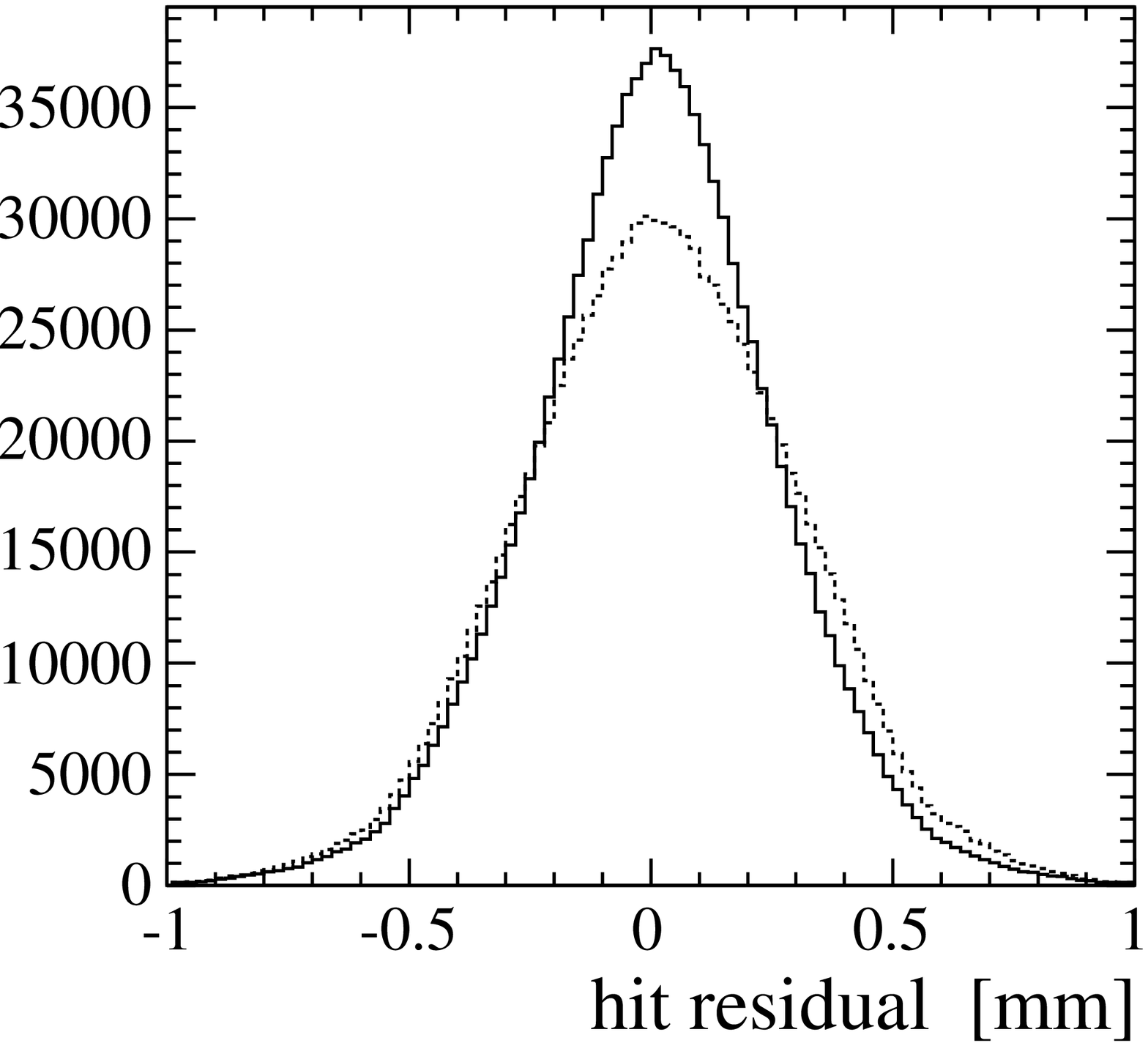}}
     \put(0,0){(a)}
     \put(45,150){LHCb OT}
     \put(230,0){(b)}
     \put(275,150){LHCb OT}
   \end{picture}    
  \caption{\small (a) Displacement of modules relative to the survey and
    (b) hit residuals in the first X-layer of station T2 before (dashed
    line) and after (continuous line) offline module alignment.}
  \label{fig:modulealignment}
\end{figure}

To achieve optimal track parameter resolution the position and
orientation of the OT modules must be known with an uncertainty that is
negligible compared to the single hit resolution. 

The OT C-frames hang on rails and can be moved outside the LHCb
acceptance to allow for maintenance work during technical stops of
the LHC. Since no survey is performed after such operations, the
reproducibility of the nominal position is important. Using track
based alignment the reproducibility has been established to be better
than 100\,\mum{}, consistent with the measurements done with the Rasnik system.

The most precise alignment information is obtained with a software
algorithm that uses charged particle trajectories~\cite{Amoraal:2012qn}. For each module and
C-frame the alignment is parametrized by three translations and three
rotations. The algorithm selects high quality tracks and subsequently
minimizes the total $\chi^2$ of those tracks with respect to the
alignment parameters. Only a subset of parameters
needs to be calibrated to obtain sufficient precision. For the
alignment of modules inside each C-frame only the translation in $x$
and the rotation in the $xy$ plane are determined. For the C-frames
themselves only the translations in $x$ and $z$ are calibrated. To
constrain redundant degrees of freedom the survey measurements are
used as constraints in the alignment procedure.

Figure~\ref{fig:modulealignment} illustrates the result of an alignment of
module positions. For this alignment, tracks were fitted
using only the OT hits. At least 18 hits per track were required. To remove
poorly constrained degrees of freedom, modules in the first $x$ and stereo ($u$)
layers of stations T1 and T3 were all fixed to their nominal position.
Figure~\ref{fig:modulealignment}(a) shows the difference between the
$x$-position of the module center relative to the survey. Statistical
uncertainties in alignment parameters are negligible and the alignment is
reproducible in data, taken under similar conditions, within about
20~\mum. Figure~\ref{fig:modulealignment}(b) shows the hit residuals in one
layer before and after alignment. A clear improvement is observed.

\begin{figure}[!t]
   \begin{picture}(250,190)(0,0)
   \put(5,0){\includegraphics[width=1.0\textwidth]{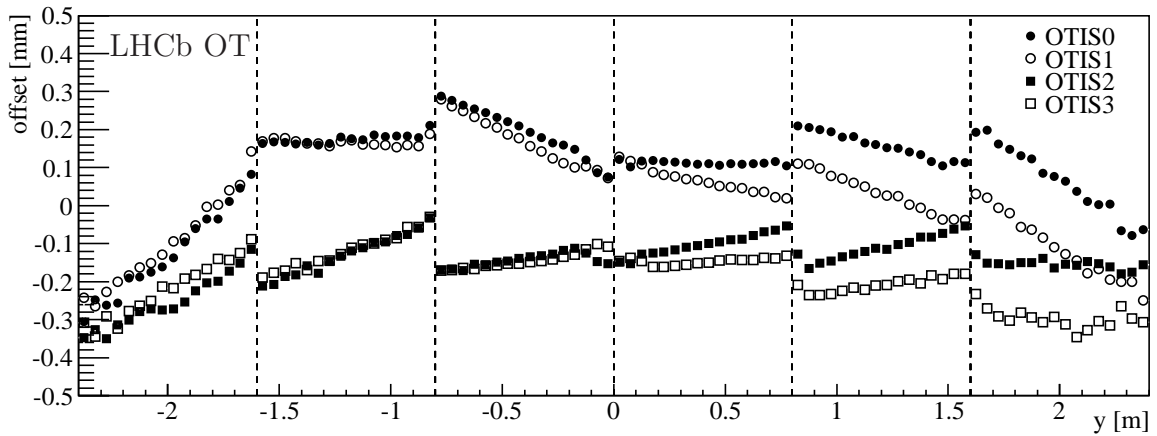}}
     \put(44,145){LHCb OT}
   \end{picture}
  \caption{\small 
    Average hit residual as function of $y$ coordinate in one particular module
    (labelled T3L3Q1M7). The four curves show residuals for the four groups of
    32 channels within one FE-module. The round markers correspond to one
    monolayer of 64 straws, whereas the square markers show the residuals of the
    second monolayer.  The vertical dashed lines indicate the position of the
    wire locators, at every 80~cm along the wire~\cite{Kozlinskiy:2013}.}
  \label{fig:residualVsHitY}
\end{figure}

The module displacements in Fig.~\ref{fig:modulealignment} are
larger than expected, based on the expected accuracy of the dowel pins that
keep the modules in place. It is assumed that the disagreement can at
least partially be explained by degrees of freedom that are not yet
corrected for, such as module deformations and the positioning of straws
within each module. Figure~\ref{fig:residualVsHitY} shows an example
of the average hit residual as a function of the coordinate along the
wire for one module. A relative displacement of the two
monolayers is observed, as well as jumps at the wire locators, 
which are placed at every 80\,cm along the wire length.
The effect on the final hit resolution is discussed in Sec.~\ref{sec:hitres}.

\label{sec:align}

\clearpage
\section{Performance}
\label{sec:performance}
The performance of the OT detector was stable in the entire first running period
of the LHC between 2010 and 2012, as was shown in the previous sections. 
No significant failures in
the LV, HV and gas systems occurred.  The details of the data quality in terms
of resolution and efficiency are described below.

\subsection{Spillover and drift-time spectrum}
\label{sec:drifttime}

In order to register all charged particle hits produced in the $pp$ interaction,
three consecutive intervals of $25\,\mathrm{ns}$ are readout upon a
positive L0-trigger. Only the first hit in the readout window is recorded,
as the first hit typically corresponds to the ionization cluster closest to the
wire, and it is thus the best estimate for the radial distance to the wire.
 
In the following, data are studied that are recorded in $75\,\mathrm{ns}$, $50\,\mathrm{ns}$
and $25\,\mathrm{ns}$ bunch-spacing data taking periods of the LHC.  These varying
conditions show the effect of so-called spillover hits on the
drift-time spectrum and straw occupancies. Distributions obtained in the
$75\,\mathrm{ns}$ bunch-crossing period are close to those observed with only
one single bunch crossing in LHCb, and therefore they will be considered free of
spillover.

\begin{figure}[b]
    \begin{picture}(250,165)(0,0)
        \put(0,  10){\includegraphics[bb=0 0 567 433, scale=0.38]{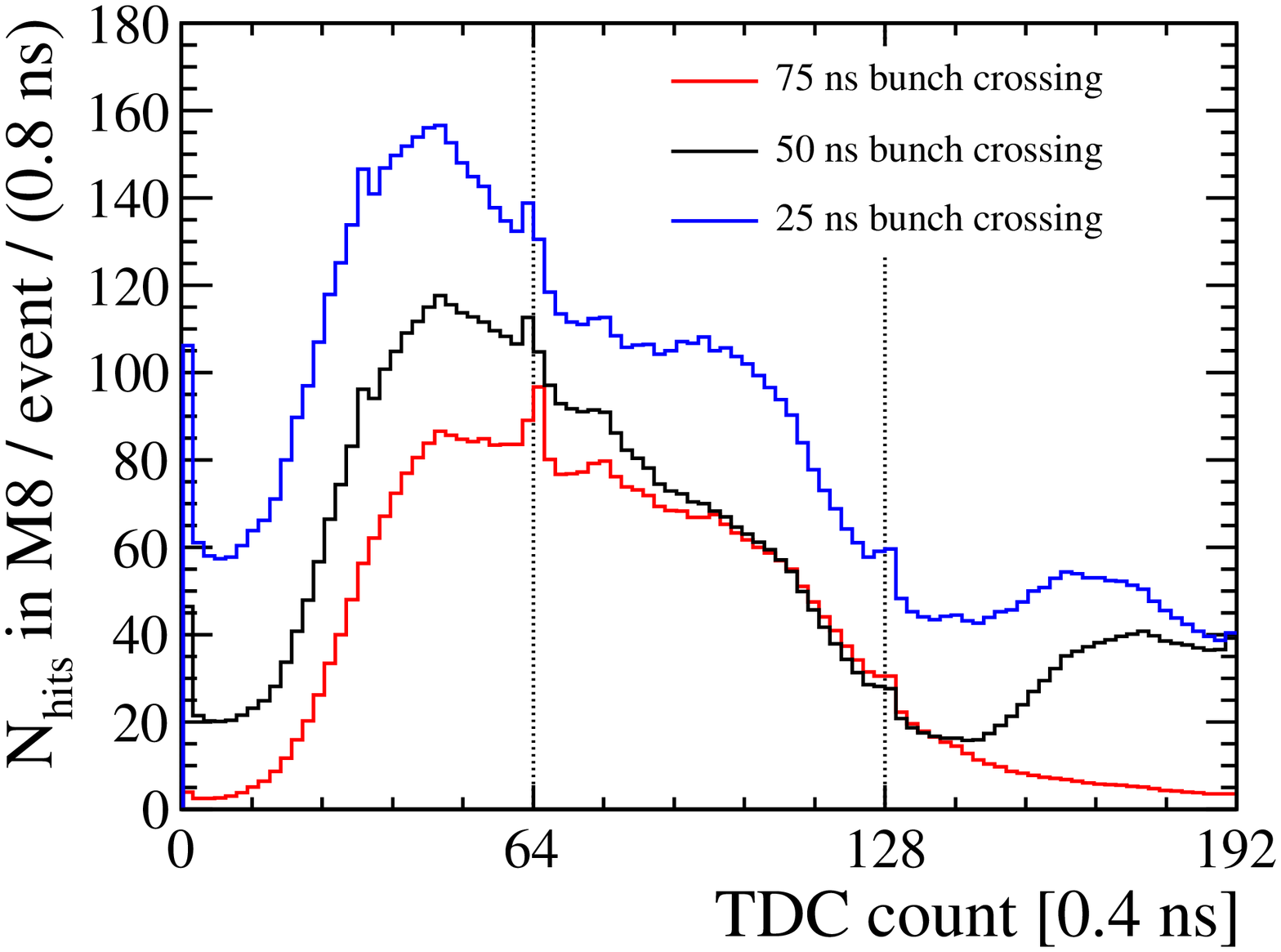}}
	\put(220,10){\includegraphics[bb=0 0 567 386,scale=0.425]{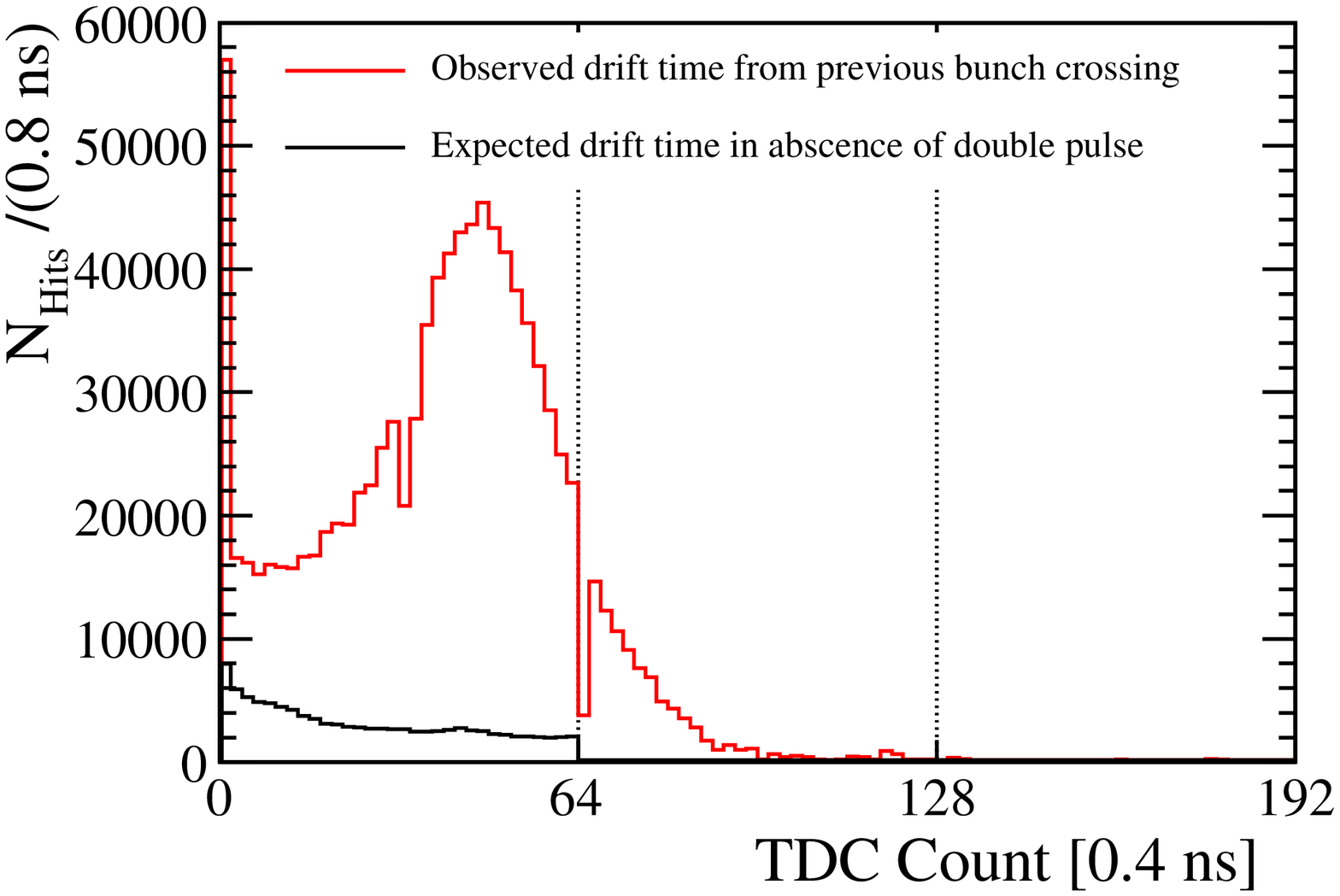}}
    \put(10,0){(a)}
    \put(148,111){LHCb OT}
    \put(240,0){(b)}
    \put(390,125){LHCb OT}
    \end{picture}
\caption{\small 
(a) Drift-time distribution in module 8, close to the beam, for
$75\,\mathrm{ns}, 50\,\mathrm{ns}, 25\,\mathrm{ns}$ bunch-crossing spacing in
red, black and blue, respectively.  The vertical lines at 64 and 128 TDC counts
correspond to 25 and 50\,ns, respectively.  The distributions correspond to all
hits in 3000 events for each bunch-crossing spacing, recorded with an average
number of overlapping events of $\mu = 1.2, 1.4$ and 1.2, for $75\,\mathrm{ns},
50\,\mathrm{ns}$ and $25\,\mathrm{ns}$ conditions, respectively.  
(b) The drift-time distribution for empty events illustrates the contribution from
spillover hits from ``busy'' previous bunch-crossings (red). The naive
expectation of the spillover distribution is shown in black, and is obtained by
shifting the nominal drift-time spectrum by $-50\,\mathrm{ns}$.}
\label{fig:drifttime}
\end{figure}

The drift-time spectrum and the occupancies presented here correspond to events
with an average number of visible $pp$ interactions per bunch crossing of about
1.4, in accordance with the typical run conditions in 2011 and 2012. The events are
triggered by any physics trigger, implying that most events contain $B$ or
$D$-decays.  The drift-time distributions for the $75\,\mathrm{ns}$,
$50\,\mathrm{ns}$ and $25\,\mathrm{ns}$ bunch-spacing conditions are shown in
Fig.~\ref{fig:drifttime}.

The typical drift-time spectrum from the (spillover-free) distribution from the
75\,ns running can be understood by inspecting Fig.~\ref{fig:clbr:tr} in
Sec.~\ref{sec:tr}.  The projection of the TR-relation results in a
linearly decreasing drift-time spectrum, assuming a flat distribution of the
distance between the tracks and the wires.  In addition, the number of earlier
hits is slightly enhanced in the drift-time distribution, since late hits are
hidden by earlier hits on the same straw, as only the first hit is recorded. The
recording of the first hit only, induces a ``digital dead-time'' starting from
the first hit until the end of the readout window at 192 TDC counts, or 75\,ns.
A second source of dead-time originates from the recovery time required by the
amplifier.  This ``analog dead-time'' lasts between 8\,ns and 20\,ns, depending
on the signal pulse height, and is usually hidden by the digital dead-time.

The black line in Fig.~\ref{fig:drifttime}(a) correspond to the data recorded in
the $50\,\mathrm{ns}$ bunch-spacing period.  The contribution from hits from the
next bunch-crossing, 50\,ns later, is visible between 128 and 192 TDC counts.
The relative contribution of these late hits from the next bunch crossing is
determined by the average occupancy in the next bunch crossing, and thus depends
on the run conditions. In principle it also depends on the occupancy of the
triggered event, and thus on the trigger configuration, but in practice that is
quite stable.  The shape of the drift-time distribution of the late spillover
hits corresponds to the nominal, spillover-free (\ie~75\,ns) drift-time
spectrum, with a shift of $+50\,\mathrm{ns}$.

The drift-time shape of the spillover hits from the previous $-50\,\mathrm{ns}$
bunch-crossing is more complex.  It contains the late hits of the drift-time
distribution from the previous bunch-crossing.  Naively, the drift-time spectrum
of these early hits can be modelled by a shift of the spillover-free
distribution by $-50\,\mathrm{ns}$, as illustrated by the black line in
Fig.~\ref{fig:drifttime}(b).  However, a traversing track can give rise to
multiple hits, which are usually not detected due to the digital ``dead-time''.
These multiple-hits, or ``double pulses'' from the previous
bunch-crossing now become visible, when they fall inside the readout window of
the triggered bunch-crossing.

In 30 to 40\% of all hits, the first arriving ionization cluster produces a
second hit that arrives about $30\,\mathrm{ns}$ later.  Several effects, such as
multiple ionizations, reflections~\cite{Guz:816183} or photon
feedback~\cite{Suvorov:2005zc}, can produce such a double pulse.  The
time-spectrum of late hits from the previous bunch-crossing, observed in the
triggered bunch crossing, is clearly isolated by studying ``empty''
bunch-crossings with ``busy'' previous bunch-crossings.  The empty and busy
bunch-crossings are selected using the total activity in the calorimeter in the
subsequent bunch-crossings.  The resulting drift-time spectrum of late hits from
busy previous bunch-crossings in empty triggered events is shown as the red line
in Fig.~\ref{fig:drifttime}(b).  The large number of double-pulses around 40 TDC
counts, or 15\,ns explains the enhancement of hits between $0$ and
$25\,\mathrm{ns}$ in the $50\,\mathrm{ns}$ bunch-crossing drift-time spectrum
compared to the spillover-free drift-time spectrum from the $75\,\mathrm{ns}$
data, see Fig.~\ref{fig:drifttime}(a).

Finally, the drift-time spectrum corresponding to the 25\,ns bunch spacing
conditions (recorded in Dec 2012) is also overlayed in
Fig.~\ref{fig:drifttime}(a).  An overall increase of the number of hits is seen
for a comparable number of overlapping events, compared to the 50 and 75\,ns
running conditions.

\subsection{Occupancy}
\label{sec:occupancy}

\begin{figure}[!t]
    \begin{picture}(250,180)(0,0)
    \put(35,190){  \includegraphics[width = 0.43\textwidth,angle=-90]{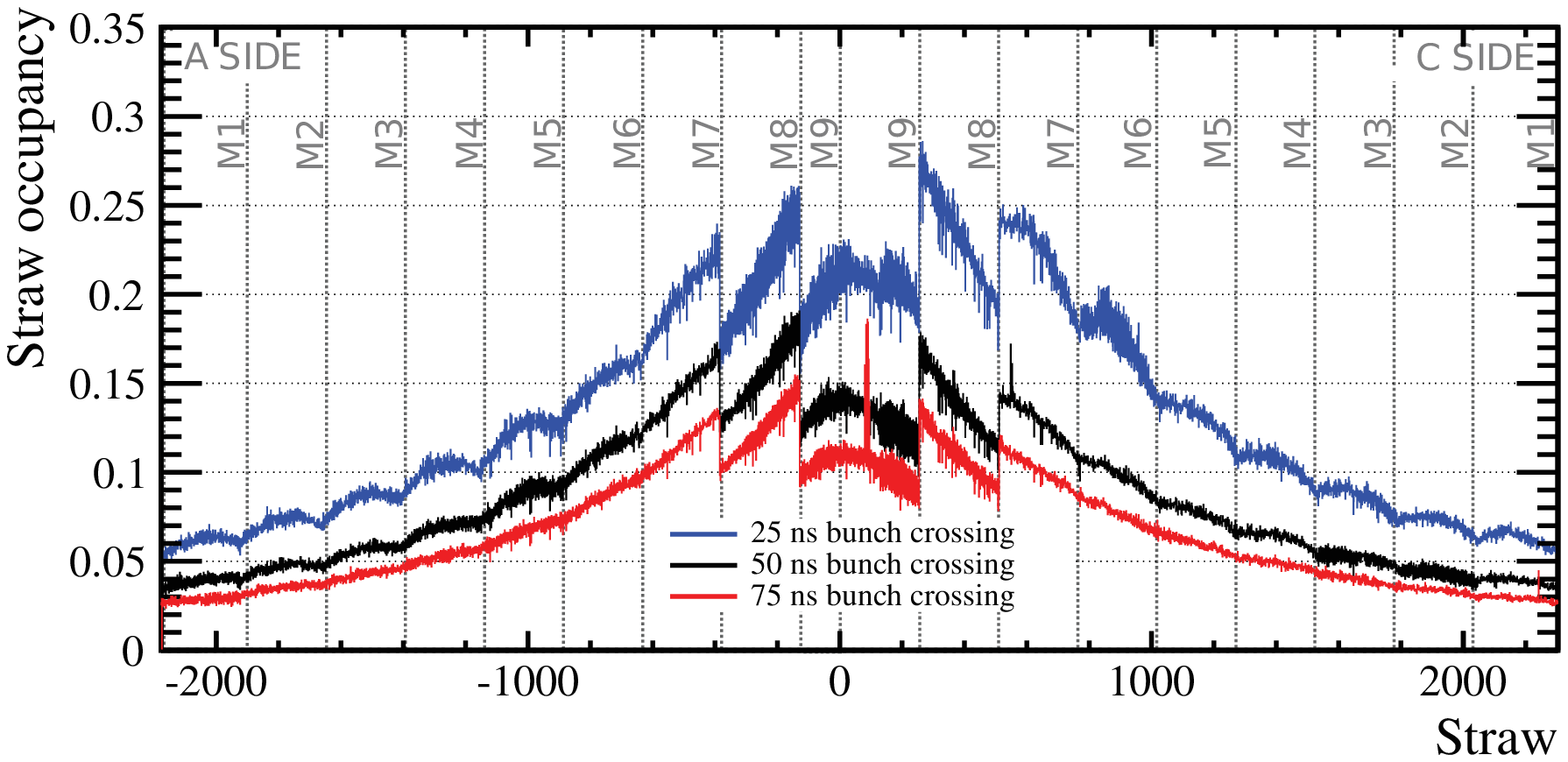}}
    \put(100,165){LHCb OT}
    \end{picture}
\caption{\small 
Straw occupancy for $75\,\mathrm{ns}, 50\,\mathrm{ns}, 25\,\mathrm{ns}$
bunch-crossing spacing in red, black and blue, respectively, for typical run
conditions with on average 1.2, 1.4 and 1.2 overlapping events per bunch
crossing, respectively. One module contains in total 256 straws, whereas the
width of one module is 340\,mm.  The steps in occupancy at the center of the
detector correspond to the location of the shorter S-modules, positioned further
from the beam in the $y$-coordinate.  The data corresponding to
$25\,\mathrm{ns}$ bunch-crossing spacing, was recorded with opposite LHCb-dipole
polarity, as compared to the other two data sets shown here.
}\label{fig:ocupmod}
\end{figure}

The occupancy per straw is shown in Fig.~\ref{fig:ocupmod} for typical run
conditions in 2011 and 2012, triggered by any physics trigger. The occupancy is shown for
events with 25, 50 and 75\,ns bunch-crossing conditions.  In absence of
spillover (\ie~ the 75\,ns case), the occupancy varies from about $15\%$ in the
innermost modules to about $3\%$ in the outermost modules. For the data taken
with $50\,\mathrm{ns}$ bunch-crossing spacing, about 30\% of all hits originate
from spillover, \ie~ from the previous bunch crossing.

\begin{figure}[!t]
    \begin{picture}(250,190)(0,0)
    \put(0,-5){ \includegraphics[width=1.00\textwidth]{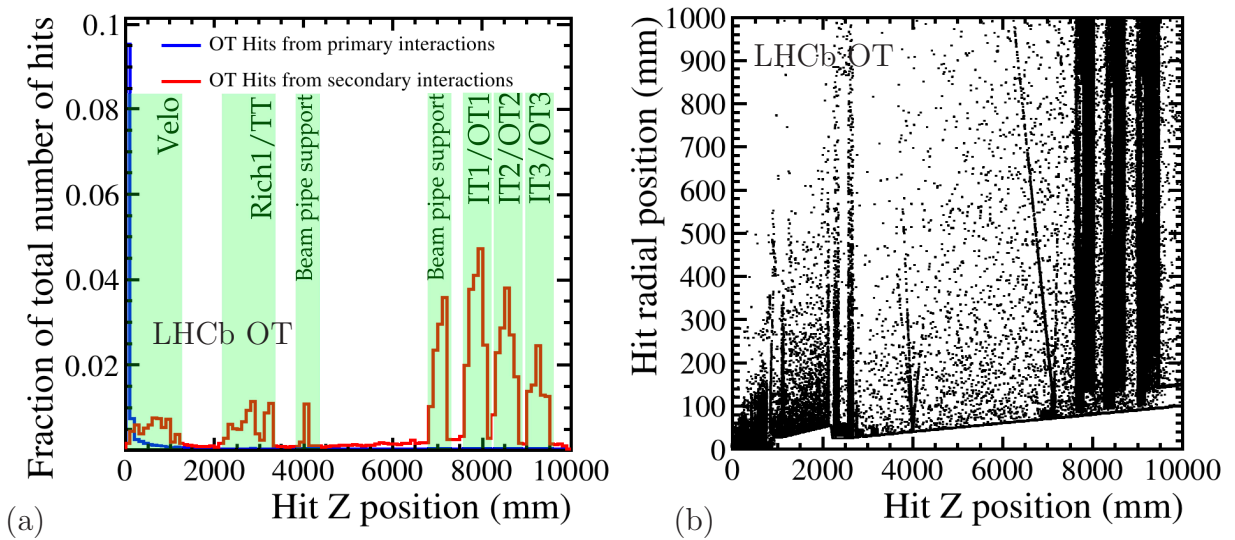}}
    \put(0,0){(a)}
    \put(250,0){(b)}
    \put(55,70){LHCb OT}
    \put(280,175){LHCb OT}
    \end{picture}
\caption{\small Coordinate of the origin of charged particles that produce a hit
in the OT detector. (a) The blue histogram peaks at $z=0$ and corresponds to
hits from particles produced at the $pp$ interaction point and their daughters,
while the hits from particles produced in secondary interactions (red)
predominantly originate from $z>0$. (b) The longitudinal and transverse position
of the origin of charged particles produced in secondary interactions, showing
the structure corresponding to the material in the detector.}
\label{fig:secondint}
\end{figure}

Monte Carlo simulations demonstrate that most of the hits originate from
secondary charged particles, produced in interactions with material. Figure
\ref{fig:secondint} shows the fraction of hits that originate from a particle
created at a given $z$ coordinate.  The hits from tracks that originate from the
genuine $pp$ interaction or a subsequent particle decay, are predominantly
located close to the interaction region. They represent $27.7\%$ (resp. $27.1\%$
and $25.7\%$) of all hits seen in station T1 (resp. T2 and T3). The remaining
hits originate from charged particles created in secondary interactions, mainly
in the support of the beam pipe situated in the magnet or in the detectors
located upstream of the detector layer (Vertex Locator, Ring Imaging Detector,
Tracker Turicensis (TT), Inner Tracker (IT) and OT).

\subsection{Hit efficiency}
\label{sec:hitefficiency}
A high single-hit efficiency is crucial, as it affects the tracking efficiency,
and eventually the physics performance of the LHCb experiment.  The efficiency
is defined as the number of observed hits in a particular detector region over
the number of expected hits in the same region.  The number of expected hits is
estimated by considering charged particle tracks in $pp$ collision data and
extrapolating the charged particle trajectory to the monolayer under study.

In order to determine the hit efficiency, good quality tracks have been
selected, requiring a $\chi^2/ndf$ (where $ndf$ are the number of degrees of
freedom) less than 2 and a minimum number of 21 hits in the OT detector.  
This corresponds to accepting about 87\% of all good tracks.
For
each track, every OT monolayer has been considered, and a hit has been searched
in the straw closest to the charged particle trajectory.  Since a track is
reconstructed by the same hits that are subsequently used for the efficiency
estimation, the large number of required hits could bias the efficiency
determination.  This has been corrected for by not considering the monolayer
under study, when counting the minimum number of hits per track.

\begin{figure}[!b]
    \begin{picture}(250,170)(0,0)
    \put(-25,5){\includegraphics[width = 0.59\textwidth]{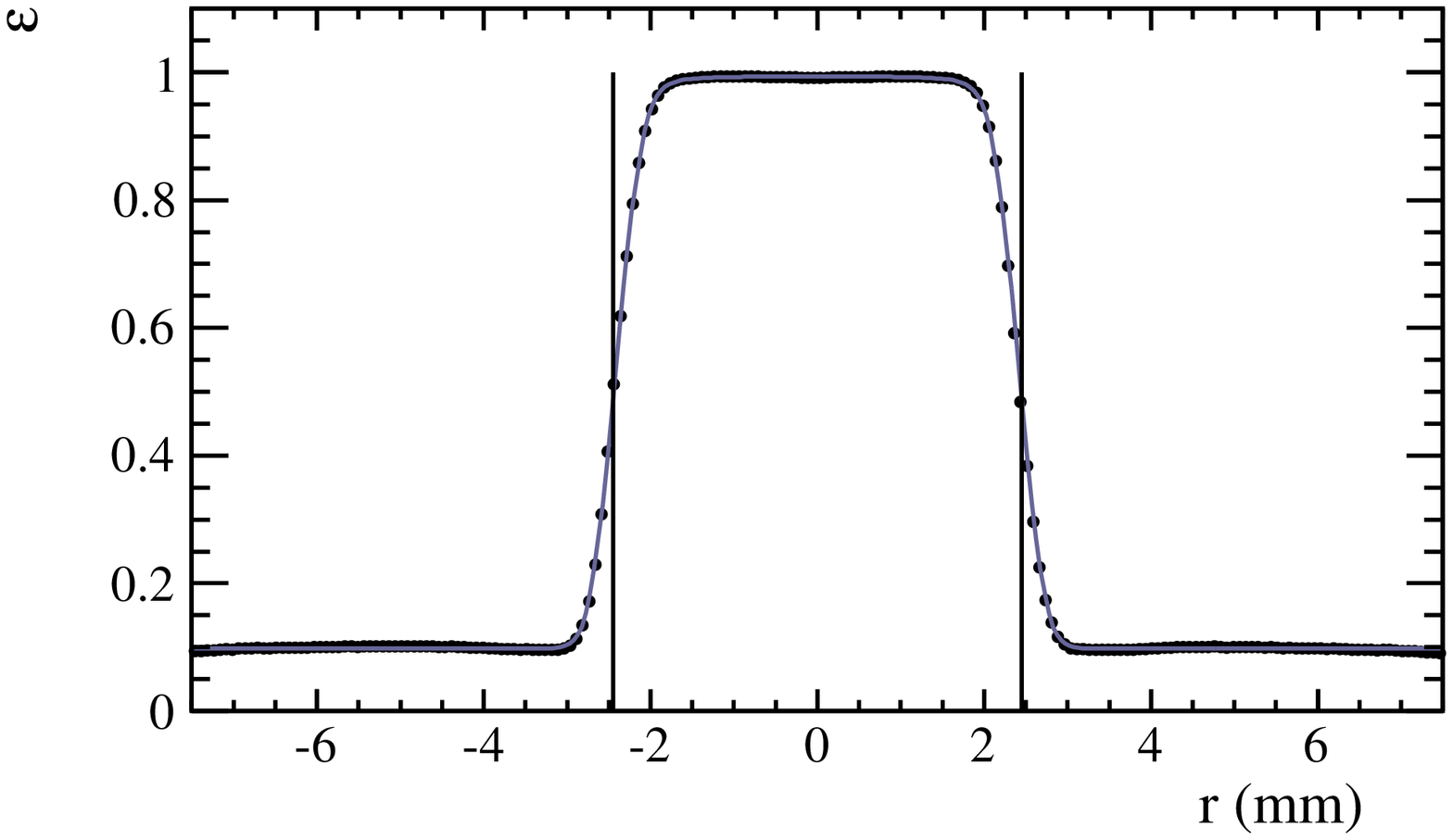}}
    \put(232,5){\includegraphics[width = 0.50\textwidth]{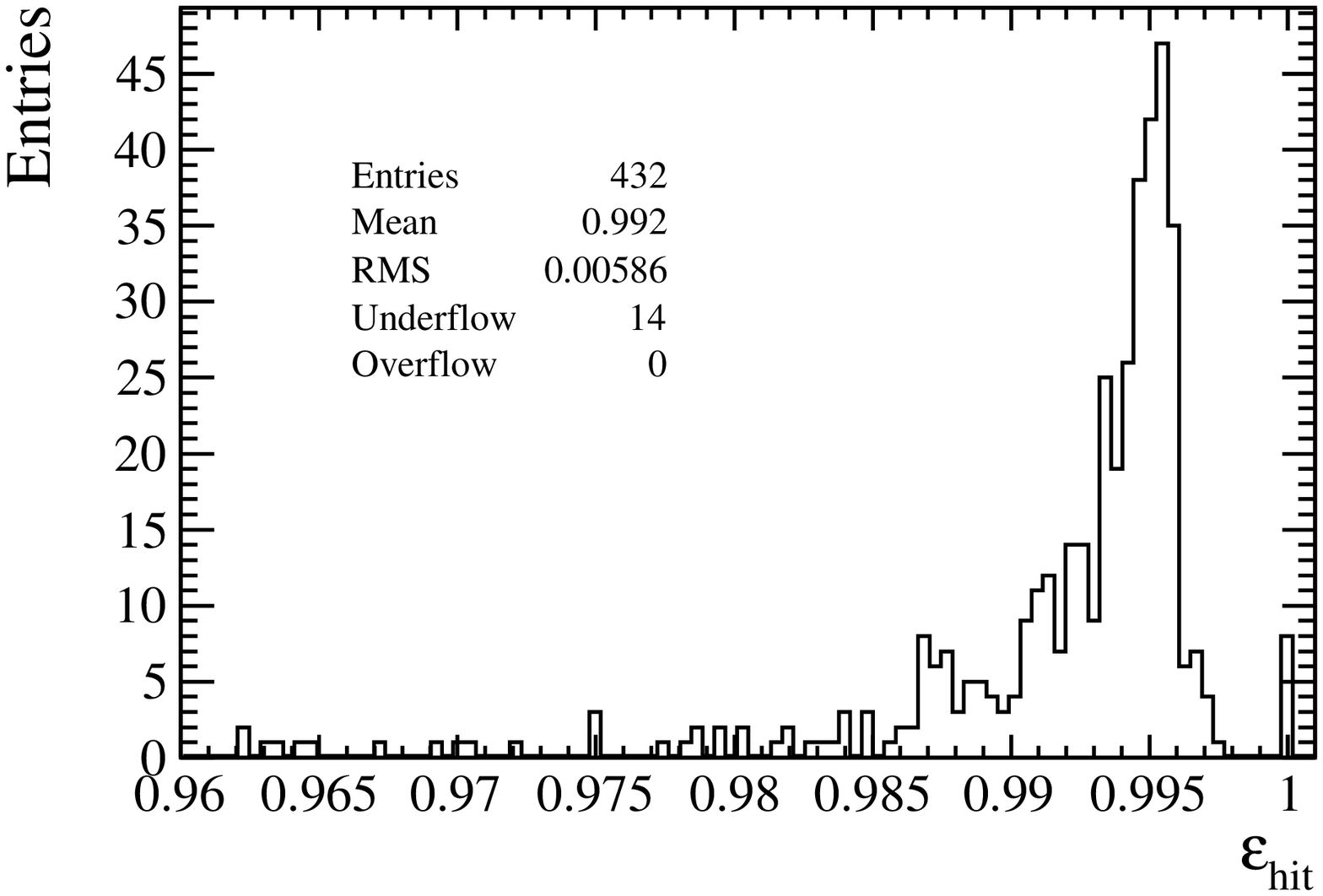}}
    \put(0,0){(a)}
    \put(250,0){(b)}
    \put(20,135){LHCb OT}
    \put(270,135){LHCb OT}
    \end{picture}
\caption{\small (a) Efficiency profile as a function of the
distance between the predicted track position and the center of the
straw, for straws in the long F-modules closest to the beampipe (module 7).
The vertical lines represent the straw tube edge at $|r|=2.45$\,mm.
(b) Histogram of the average efficiencies per half module (128 channels),
at the center of the straw, $|r|<1.25$\,mm, for runs 96753, 96763 and 96768 on 22 July 2011.
}\label{fig:profile}
\end{figure}

The hit efficiency is studied as a function of the distance between the
predicted track position and the center of the considered straw.  The resulting
single-hit-efficiency profile is shown in Fig.~\ref{fig:profile}(a), summed for
all straws in the long modules closest to the beam-pipe (module 7).

The shape of the efficiency profile can be understood by considering two
effects. Near the straw tube edge, the path length of ionizing particles inside
the gas volume is limited, resulting in a sizeable probability for not ionizing
the gas.  This can be described with a Poissonian distribution for the
single-hit probability.  Secondly, the finite track resolution smears the
distribution at the edge of the straw tube, lowering the efficiency inside and
increasing the efficiency outside of the straw.  The finite probability to
detect a hit outside straw tube originates from random hits unrelated to the
track under study, and is proportional to the average occupancy in that part of
the detector.

The straw tube profile can thus be fitted with the following line shape, which
describes the efficiency as a function of the distance $r$ from the center of
the straw,

\begin{eqnarray}\label{eq:efficiency}
 p(r)           & = & 
1 - \Big( 1 - \varepsilon(r) \otimes \rm{Gauss}(r|0, \sigma) \Big) 
\cdot (1- \omega),\\ 
\mathrm{with} \,\,\,\,\,\, \varepsilon(r)  & = & 
\varepsilon_0 \left( 1 - e^{\frac{-2 \sqrt{R^2 - r^2}}{\lambda}}  \right),
\nonumber
\end{eqnarray}
where $R=2.45$~mm is the inner radius of the straw, $\omega$ is the average
occupancy, $\lambda$ is the effective ionization length of a charged particle in
the gas volume, and $\sigma$ is the track resolution.

The deviation from the perfect efficiency is quantified in
Eq.~\eqref{eq:efficiency} by $\varepsilon_0$.  However, in the following, an
operationally straightforward definition of the single-hit efficiency is used.
The single-hit efficiency per straw is defined as the average hit efficiency
$\varepsilon_{hit}$ in the limited range close to the wire, $|r|<1.25$\,mm.  The
inefficient regions between two straws lead to the maximum efficiency of 93\%,
integrated over the monolayer, and is calculated by taking the ratio of the
straw diameter of 4.9\,mm over the pitch, 5.25\,mm.  The inefficient regions are
covered by the neighbouring monolayer in the same module, which is staggered by
half a straw pitch.

\begin{table}[!b]
\caption{\small Average single-hit efficiencies $\varepsilon_{hit}$ near the
center of the straws, $|r|<1.25$\,mm,
for different module positions of the OT detector.}\label{tab:module_efficiency}
\begin{center}
\begin{tabular}{cl}
\hline
Module position& Efficiency (\%)\\
\hline
1              &  $98.085 \pm 0.011$ \\
2              &  $99.130 \pm 0.005$ \\
3              &  $99.279 \pm 0.003$ \\
4              &  $99.277 \pm 0.003$ \\
5              &  $99.282 \pm 0.002$ \\
6              &  $99.342 \pm 0.002$ \\
7              &  $99.286 \pm 0.002$ \\
8              &  $99.200 \pm 0.002$ \\
9              &  $99.351 \pm 0.003$ \\
\hline
\end{tabular}
\end{center}
\end{table}

A fit to the straw efficiency profile, using Eq.~\eqref{eq:efficiency}, 
separately for the profiles of the nine module positions, 
yields the following average parameters,
$\langle \lambda \rangle = 0.79 \pm 0.09$~mm, 
$\langle \varepsilon_0 \rangle = 0.993 \pm 0.003$, 
$\langle \sigma \rangle = 0.26 \pm 0.06$~mm, 
$\langle \omega \rangle = 0.07 \pm 0.02$,
where the quoted uncertainty is the standard deviation from the values obtained
for the nine different module positions.  
Note that these parameters are averaged over the different module positions,
corresponding to different conditions.  For
example, the measured occupancy $\omega$ varies from $4.7\%$ to $9.7\%$
depending on the distance of the module to the beam.  Fits to the single-hit
efficiency profiles show that the efficiency is close to maximal, \ie~ the value
for $\varepsilon_0$ is consistent with unity.

For each half-module, corresponding to one FE box, the average single-hit
efficiency $\varepsilon_{hit}$ has been calculated and the result is shown in
Fig.~\ref{fig:profile}(b).  The efficiency distribution peaks around 99.5\%,
consistent with the measurements from beam tests~\cite{vanApeldoorn:2005ss}.
The average of the distribution is about 99.2\%.  This value is consistent with
the fit to the straw tube profile, shown in Fig.~\ref{fig:profile}(a).  The
large hit efficiency is a prerequisite for large efficiency to reconstruct
charged particle's tracks in LHCb.  The average tracking efficiency is approximately 95\%
in the region covered by the LHCb detector~\cite{LHCb-DP-2013-002}.

The modules that are located at the edge of the geometrical acceptance of
LHCb, in particular the outermost modules in the first station, detect a relatively
small number of tracks. 
All eight FE boxes in Fig.~\ref{fig:profile}(b) with a value of the efficiency
exactly equal to 1, and 11 out of the 14 FE boxes with $\varepsilon_{hit}<96\%$,
are attached to modules located most distant from the beampipe (module 1), and suffer from
few tracks in the efficiency determination.  The remaining three FE boxes with
$\varepsilon_{hit}<96\%$ suffer from hardware problems, representing $3/432 =
0.7\%$ of all FE boxes.

In order to calculate the average efficiency for each module position, modules
with few tracks, ie. with an efficiency lower than 96\%, or with an 
efficiency equal to unity,
have been discarded.  The average efficiency thus obtained for each module
position is listed in Table~\ref{tab:module_efficiency} where the reported
uncertainties are statistical.

As shown above, the decrease of the hit efficiency close to straw edge is
partially due to the fact that the charged particle traverses a short distance
through the straw volume.  Hence, the probability to not form an ionization
cluster increases towards the straw edge.  Alternatively, the effective
ionization length $\lambda$ can be probed by selecting only those tracks that
pass {\em close} to the wire.  In contrast to the first method exploiting
Eq.~\ref{eq:efficiency}, here the determination of the ionization length is {\em not}
affected by absorption of drifting electrons.  The larger the ionization length,
the more hits will exhibit a {\em large} drift-time, as the ionization does not
necessarily occur close to the wire.
The effective ionization length $\lambda$ extracted from particles traversing
the straw within $|r|<0.1$~mm amounts to about 0.7~mm~\cite{Kozlinskiy:2013,Tuning:2013},
consistent with $0.79 \pm 0.09$~mm, as obtained above.

\subsection{Hit resolution}
\label{sec:hitres}
\begin{figure}[!b]
    \begin{picture}(250,175)(0,0)
      \put(0,  5){\includegraphics[width=0.5\textwidth]{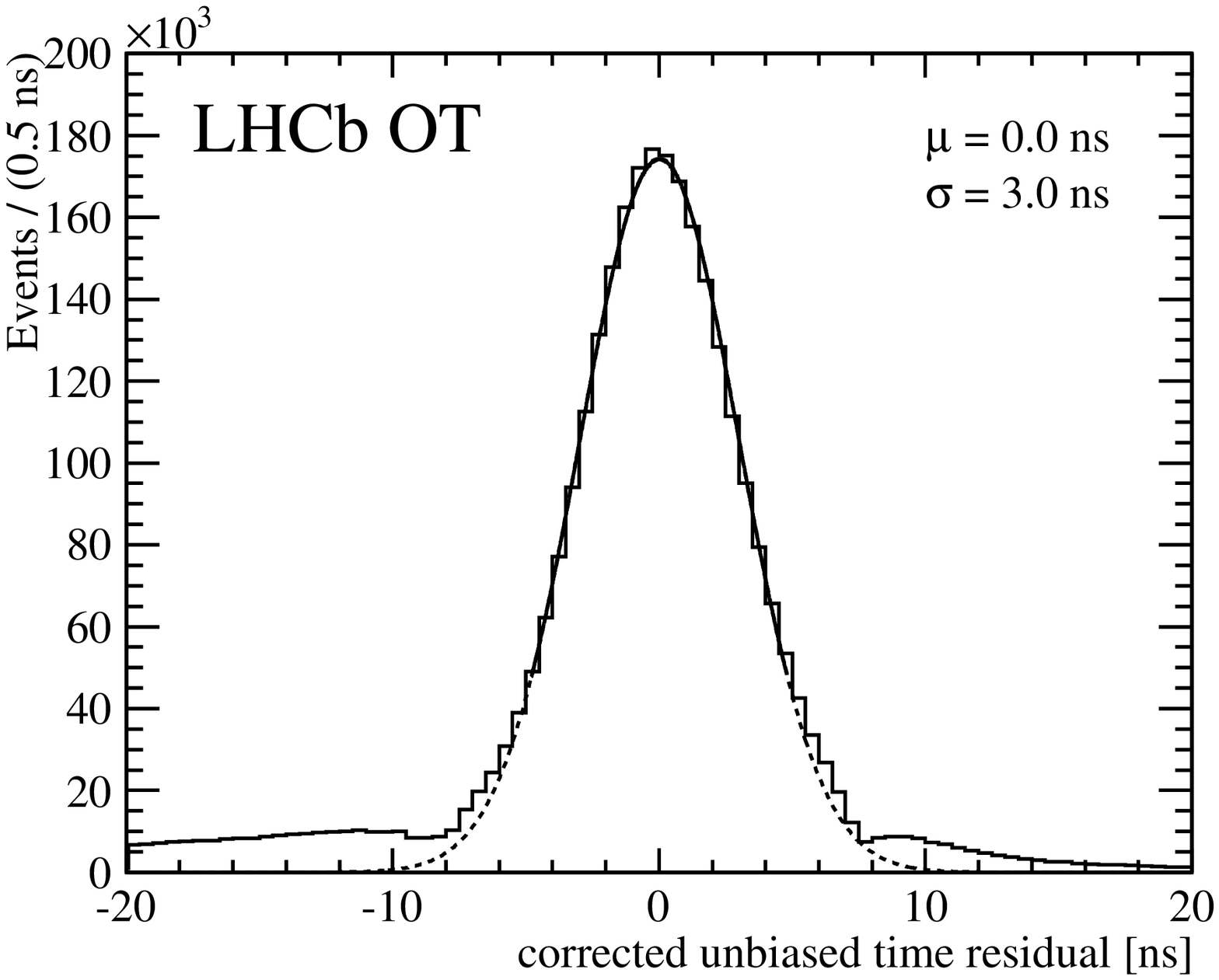}}
      \put(230,5){\includegraphics[width=0.5\textwidth]{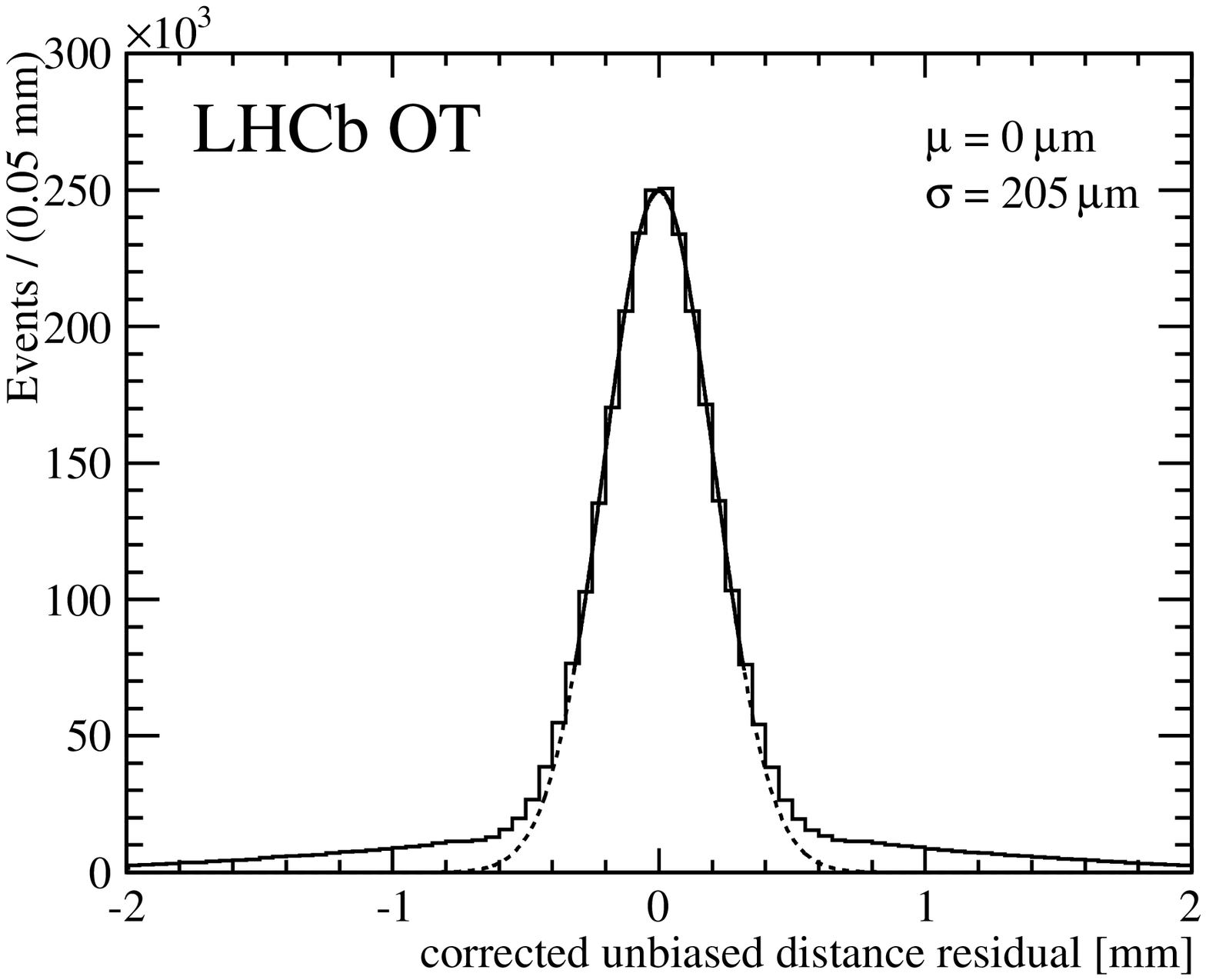}}
      \put(0,0){(a)}
      \put(230,0){(b)}
      \end{picture}
\caption[Single hit resolution]{\small
  \label{fig:resolution}
  (a) Drift-time residual distribution and 
  (b) hit distance residual distribution~\cite{Kozlinskiy:2013}.
  The core of the distributions (within $\pm1 \sigma$) are fitted with a Gaussian function
  and the result is indicated in the figures.
}
\end{figure}

The single hit resolution is determined using good quality tracks, selected by
requiring a momentum larger than 10\,GeV, at least 16 OT hits and a track-fit
$\chi^2/ndf<2$ (excluding the hit under study, and excluding any hit in the
neighbouring monolayer in the same module).  For a given track, the drift-time
and the hit position in a straw are predicted, and compared with the measured
drift-time and position, respectively.  The resulting distribution of the
drift-time residuals and hit position residuals are shown in
Fig.~\ref{fig:resolution}.  

\begin{figure}[!t]
    \begin{picture}(250,170)(0,0)
      \put(0,  5){\includegraphics[width=0.5\textwidth]{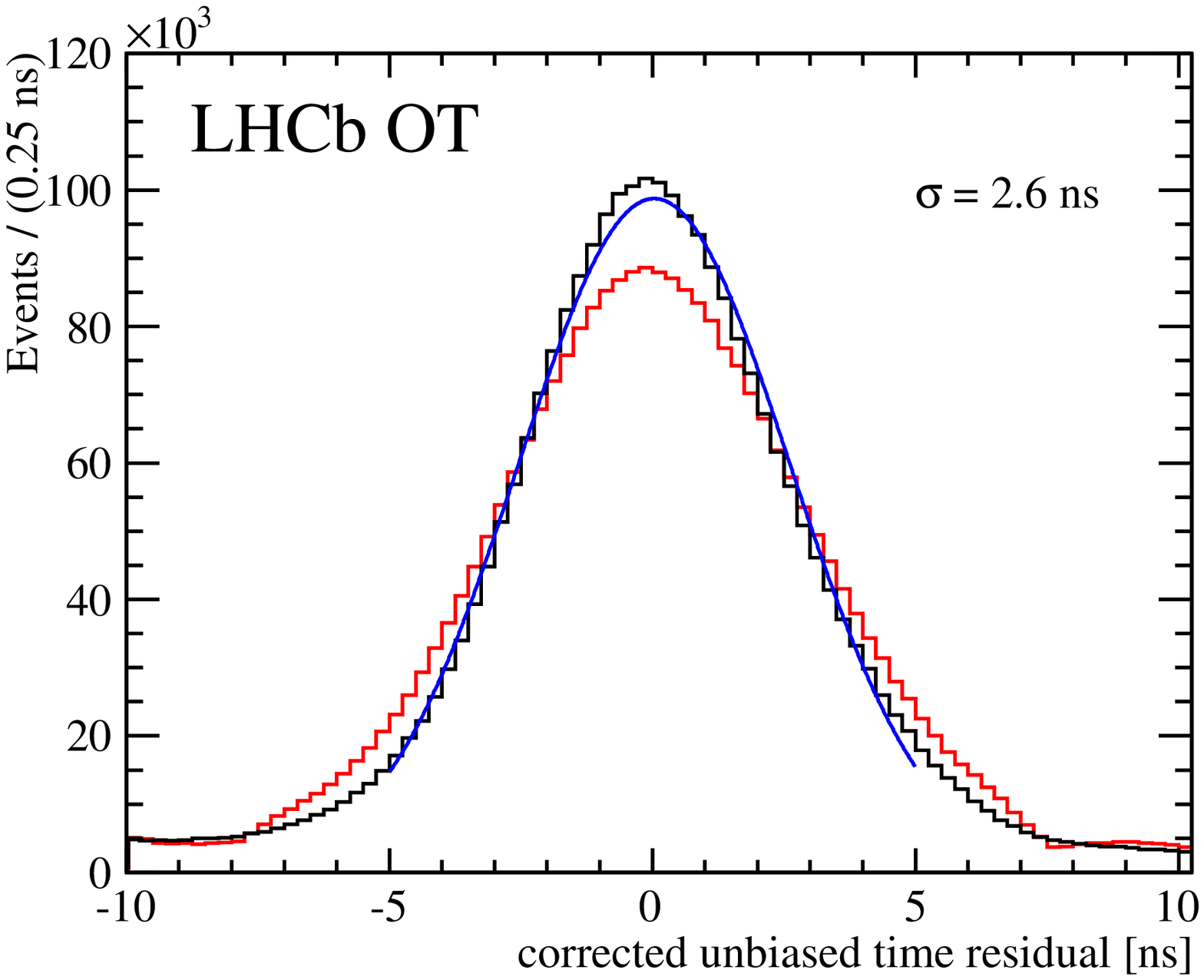}}
      \put(230,5){\includegraphics[width=0.5\textwidth]{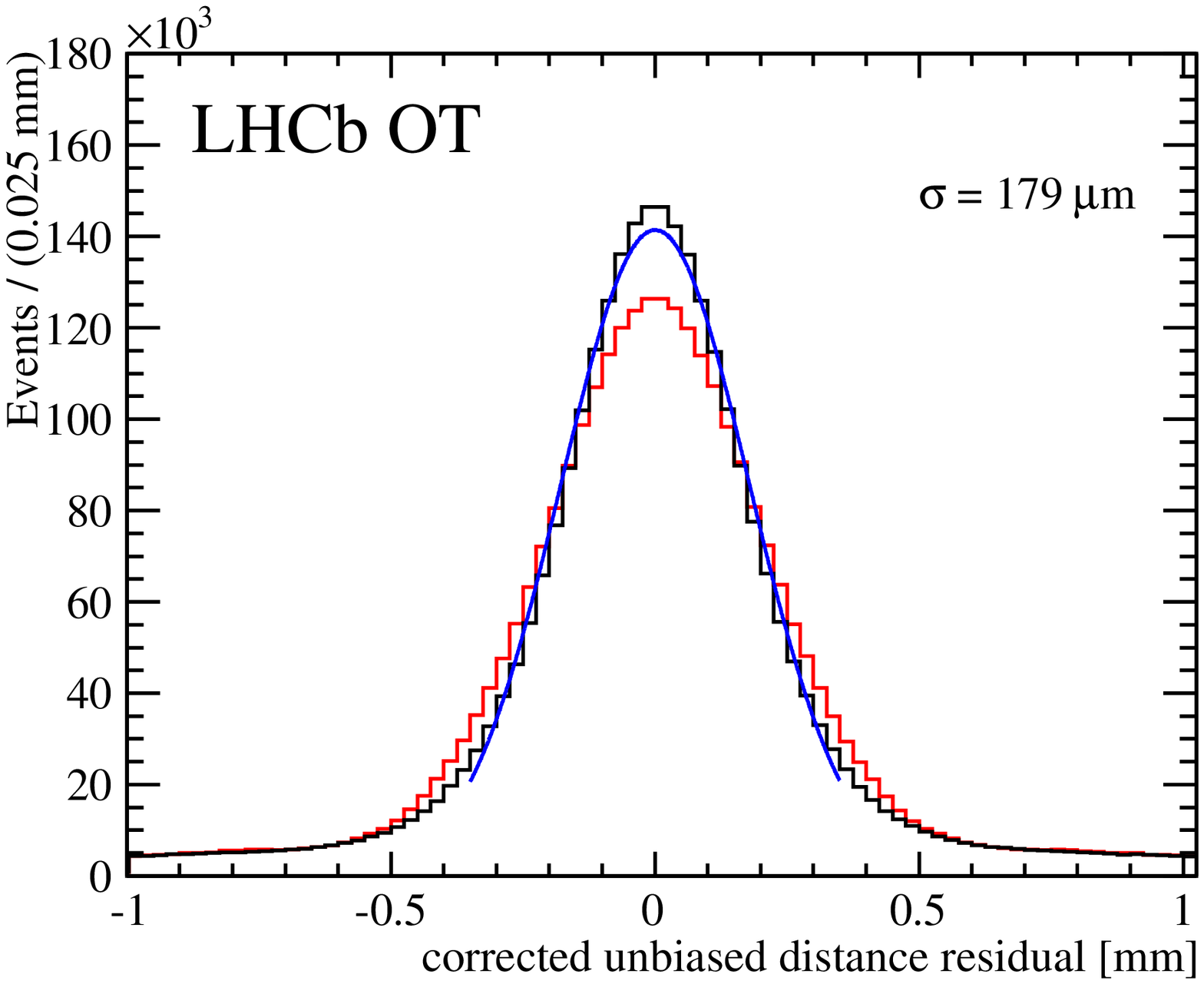}}
      \put(0,0){(a)}
      \put(230,0){(b)}
      \end{picture}
\caption[Single hit resolution]{\small 
  \label{fig:resolution2}
  Improvement in 
  (a) drift-time residual distribution and 
  (b) hit distance residual distribution, (red) before and (blue) after 
  allowing for a different horizontal displacement per half monolayer, 
  corresponding to 64 straws~\cite{Kozlinskiy:2013}.
}
\end{figure}

The drift-time residual distribution has a width of 3\,ns which is dominated by
the ionization and drift properties in the counting gas. The granularity of the
step size of the TDC of 0.4\,ns has a negligible impact on the drift-time
resolution.  The hits in the left tail of the drift-time residual distribution
are early hits, that do not originate from the track under study, but instead
are a combination of noise hits, hits from different tracks in the same bunch
crossing, and hits from tracks from previous bunch crossings (spill-over hits).

The hit distance residual distribution has a width of about 205\,$\mu$m, which
is close to the design value of 200~$\mu$m. An improvement of the hit position
resolution is foreseen when the two monolayers within one detector module are
allowed to be relatively displaced to each other in the global LHCb alignment procedure.  By
allowing a different average horizontal displacement per half monolayer,
containing 64 straws, a single hit resolution of approximately 180\,$\mu$m is in
reach, see Fig.~\ref{fig:resolution2}. Also allowing for a rotation of each half
monolayer, improves the single hit resolution further to 160 \,$\mu$m.  These
values refer to a Gaussian width of the resolution, determined from a fit to the
residual distribution, within two standard deviations of the mean.  This is in
good agreement with the hit resolution below 200\,$\mu$m, as obtained in beam
tests~\cite{vanApeldoorn:2005ss}.

\subsection{Monitoring of faulty channels}
\label{sec:badch}
\begin{figure}[!t]
    \begin{picture}(250,160)(0,0)
        \put(25,0){\includegraphics[scale=0.65]{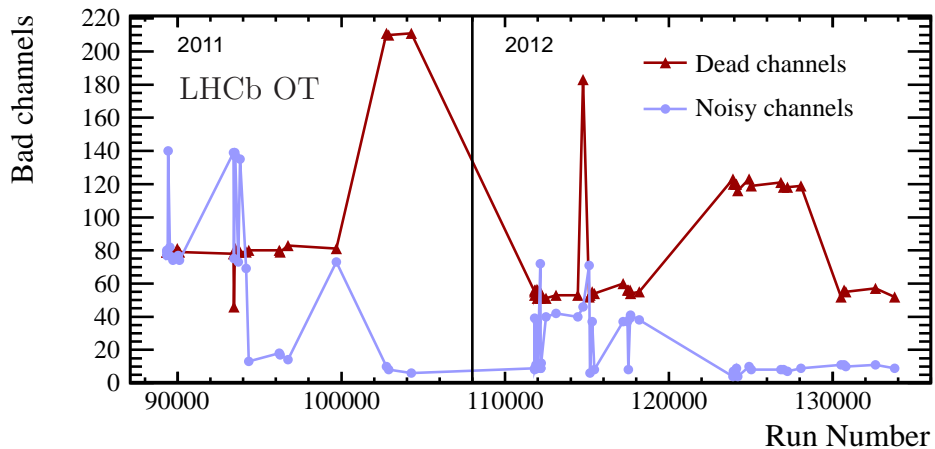}}
    \put(95,135){LHCb OT}
    \end{picture}
\caption{\small The evolution of number of dead and noisy channels as
function of run number in the 2011 and 2012 running periods.
The definition of dead and noisy channels is given in the text.
The three periods with larger number of dead channels, correspond to 
periods with a problem affecting one entire front-end box.}
\label{fig:badch}
\end{figure}

Noisy or dead channels due to malfunctioning front-end electronics are timely
identified through the analysis of the calibration runs as described in
Sec.~\ref{sec:commissioning}.  With the full offline data set available, the
performance of individual channels is also monitored by comparing the occupancy
to the expected value.  First, the performance of entire groups of 32 channels
is verified.  Then, within a group of 32 channels, the occupancy is compared to
the truncated mean, after correcting for the dependence of the occupancy on the
distance to the beam.  If the occupancy is above (below) 6 standard deviations
from the truncated mean, the channel is declared ``noisy'' (``dead'').  For a
typical run recorded at the end of 2012 (run 133785), when all front-end modules were
functioning properly, the OT contained 52 dead channels and 8 noisy channels,
evenly distributed over the detector.
The evolution of the number of bad channels throughout the 2011 and 2012 
running periods is shown in Fig.~\ref{fig:badch}.~\footnote{%
The three periods with larger number of dead channels correspond to 
a broken laser diode (VCSEL) between September and December 2011 at location T1L3Q0M2,
a broken fuse in May 2012 at location T3L3Q0M8, 
and desynchronization problems between July and September 2012 at location T2L2Q0M9.
Note that the front-end box at location M9 on the C-side reads out only 64 straws.}

\subsection{Radiation tolerance}
\label{sec:ageing}
It was discovered that, in contrast to the excellent results of extensive ageing
tests in the R\&D phase, final production modules suffered from gain loss after
moderate irradiation (i.e. moderate collected charge per unit time) in
laboratory conditions. The origin of the gain loss was traced to the formation
of an insulating layer on the anode wire~\cite{Bachmann:2010zz}, that contains
carbon and is caused by outgassing inside the gas volume of the plastifier
contained in the glue~\cite{Tuning:2011zzb}. Remarkably, the gain loss was only
observed upstream of the source position with respect to the gas flow.

A negative correlation was observed between the ageing rate and the production
of ozone~\cite{Bachmann:2010zz}, which suggests that the gain loss is prevented
under and downstream of the source due to the formation of ozone in the
avalanche region.  As a consequence it was decided to add 1.5\% O$_2$ to the
original gas mixture of Ar/CO$_2$, to mitigate possible gain loss.  In addition,
a beneficial effect from large induced currents was observed, which removed the
insulating layers from irradiation in the laboratory.  These large currents can
either be invoked by large values of the high voltage in the discharge regime
(dark currents), or by irradiating the detector with a radioactive
source~\cite{Tuning:2011zzb}.

\begin{figure}[!t]
  \begin{center}
    \begin{picture}(400,160)(0,0)
    \put(-35,10){\includegraphics[scale=0.42]{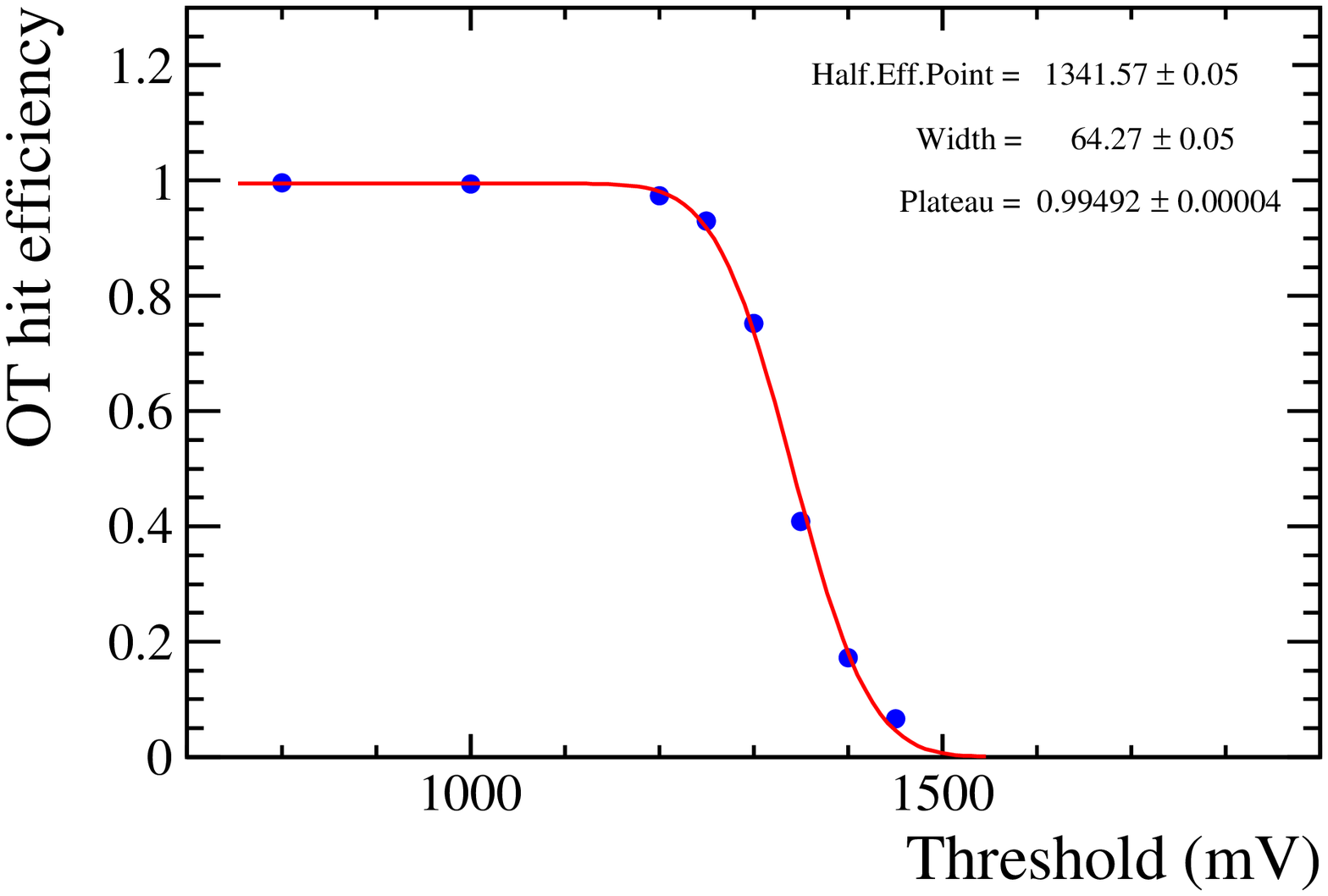}}
    \put(200,10){\includegraphics[scale=0.42]{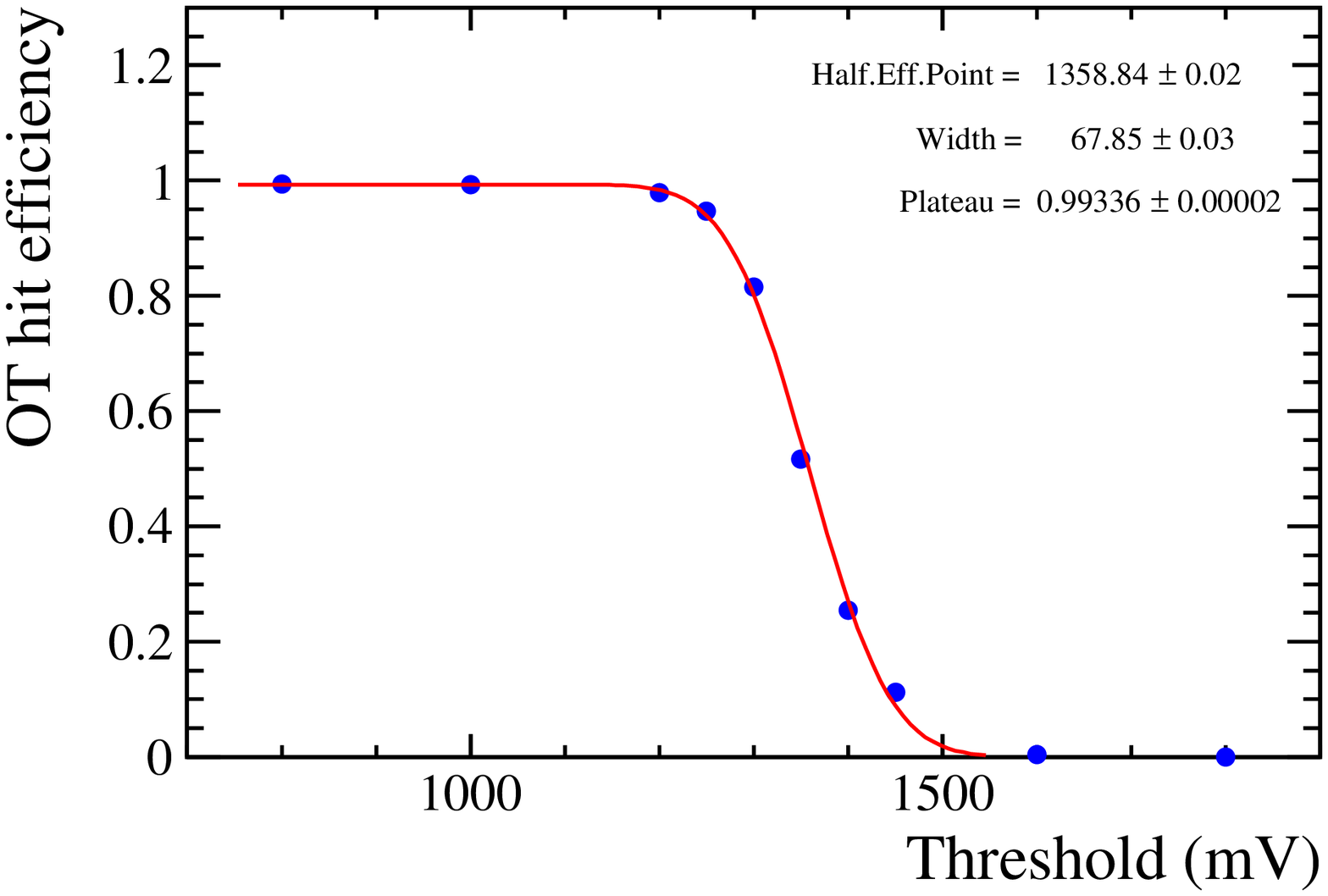}}
    \put(-10, 0){(a)}
    \put(210,0){(b)}
    \put(130,105){LHCb OT}
    \put(130, 90){Aug 2010}
    \put(365,105){LHCb OT}
    \put(365, 90){Dec 2012}
    \end{picture}
      \caption[S-curves for inner region and all layers]{\small 
      Hit efficiency as a function of amplifier threshold in (a) August
      2010 and (b) December 2012 for the inner region, defined as $\pm
      60\,\mathrm{cm}$ in $x$ and $\pm 60\,\mathrm{cm}$ in $y$ from the central
      beam pipe, summed over all OT layers. 
      Note that the threshold value of 1350\,mV, where the efficiency
      is 50\%, is much higher than the operational threshold of 800\,mV, and is
      equivalent to multiple times the corresponding average hit charge.
      }
      \label{fig:scurves}
  \end{center}
\end{figure}

No signs of gain loss have been observed in the 2010 to 2012 data taking period
of LHCb, corresponding to a total delivered luminosity of 3.5\,fb$^{-1}$.  Most
of the luminosity was recorded in 2011 and 2012, corresponding to about
$10^7$\,s of running at an average instantaneous luminosity of $3.5\times
10^{32}$\,cm$^{-2}$s$^{-1}$, and the region closest to the beam accumulated an
integrated dose equivalent to a collected charge of 0.12\,C/cm.  Possible
changes in the gain are studied by increasing the amplifier threshold value
during LHC operation, and comparing the value where the hit efficiency drops,
see Fig.~\ref{fig:scurves}.  This value of the amplifier threshold can be
converted to hit charge, which provides information on the change of the detector gain.
This method to measure the gain variations is outlined in detail in
Ref.~\cite{vanEijk:2012dx}.

\section{Conclusions}
\label{sec:conclusions}

The Outer Tracker has been operating in the 2010, 2011 and 2012 running periods
of the LHC without significant hardware failures. The low voltage, high voltage
and gas systems showed a reliable and stable performance.  Typically 250
channels out of a total of 53,760 channels were malfunctioning, resulting in
99.5\% working channels.  The missing channels were mainly caused by problems in
the readout electronics, whereas only a handful channels could not stand the
high voltage on the detector.

The occupancy of the Outer Tracker detector of typically 10\% was larger than
anticipated, due to twice larger instantaneous luminosity at LHCb with half the
number of bunches in the LHC, compared to the design specifications.  Despite
these challenging conditions, the Outer Tracker showed an excellent performance
with a single-hit efficiency of about 99.2\% near the center of the straw, and a
single hit resolution of about 200\,$\mu$m.  No signs of irradiation damage have
been observed.

\section*{Acknowledgements}
\noindent 
We wish to thank our colleagues of the CERN Gas Group for their continuous
support of the Outer Tracker gas system.
We also express our gratitude to our colleagues in the CERN
accelerator departments for the excellent performance of the LHC. We
thank the technical and administrative staff at the LHCb
institutes. We acknowledge support from CERN and from the national
agencies: CAPES, CNPq, FAPERJ and FINEP (Brazil); NSFC (China);
CNRS/IN2P3 and Region Auvergne (France); BMBF, DFG, HGF and MPG
(Germany); SFI (Ireland); INFN (Italy); FOM and NWO (The Netherlands);
SCSR (Poland); MEN/IFA (Romania); MinES, Rosatom, RFBR and NRC
``Kurchatov Institute'' (Russia); MinECo, XuntaGal and GENCAT (Spain);
SNSF and SER (Switzerland); NAS Ukraine (Ukraine); STFC (United
Kingdom); NSF (USA). We also acknowledge the support received from the
ERC under FP7. The Tier1 computing centres are supported by IN2P3
(France), KIT and BMBF (Germany), INFN (Italy), NWO and SURF (The
Netherlands), PIC (Spain), GridPP (United Kingdom). We are thankful
for the computing resources put at our disposal by Yandex LLC
(Russia), as well as to the communities behind the multiple open
source software packages that we depend on.